\documentclass[aps,pra,twocolumn,showpacs,superscriptaddress]{revtex4-1}
\usepackage{graphicx}
\usepackage{dcolumn}
\usepackage{bm}
\usepackage{color}
\usepackage{times}
\usepackage{amssymb}
\usepackage{amsmath}
\usepackage{epsfig}
\usepackage{epstopdf}
\usepackage{dsfont}
\usepackage{subfigure}
\usepackage[colorlinks=true,citecolor=blue, linkcolor=blue,urlcolor=blue]{hyperref}
\usepackage{setspace}
\usepackage{bookmark}
\usepackage{CJKutf8}
\usepackage{amsmath}
\usepackage{multirow}
\usepackage{booktabs} %
\usepackage{ulem} 
\usepackage{xcolor}

\definecolor{white}{rgb}{1,1,1}
\pagecolor{white}
\begin{document}
		
		\title{Robust composite two-qubit gates for silicon-based spin qubits}
		
		\author{Yang-Yang Yu}
		\email{These authors contributed equally to this work.}
		\affiliation{School of Physics, Anhui University, Hefei 230601, China}
		\author{Guang-Hui Zhang}
		\email{These authors contributed equally to this work.}
		\affiliation{School of Physics, Anhui University, Hefei 230601, China}
		\author{Yan-Jie He}
		\affiliation{School of Physics, Anhui University, Hefei 230601, China}
		\author{Jun Wu}
		\affiliation{School of Physics, Anhui University, Hefei 230601, China}
		\author{Xue-Ke Song}
		\email{songxk@ahu.edu.cn}
		\affiliation{School of Physics, Anhui University, Hefei 230601, China}
		\author{Dong Wang}
		\affiliation{School of Physics, Anhui University, Hefei 230601, China}
		
		\date{\today}
		
		\begin{abstract}
		We propose a universal approach based on Hamiltonian inverse engineering to realize a set of parameterized two-qubit gates. 
		This method possesses unique advantages to simultaneous control of transitions among four energy levels, providing a simpler and effective way to construct composite two-qubit gates with fewer operations than traditional methods. Applied to silicon double quantum dots (DQDs), one can realize a one-step fSim gate and a B gate with only one pulse switch. More importantly, the method can be further integrated with various optimization theories to enhance gate performance. Based on quantum optimal control theory, we develop a high-fidelity fSim gate scheme with experimentally feasible pulse shapes, featuring an average gate time of 50 ns and a theoretical fidelity of 99.95\% in the presence of decoherence and approximation error. By incorporating geometric quantum gate principles, we propose a combined geometric and dynamic fSim gate scheme. Numerical simulations demonstrate that this hybrid scheme exhibits stronger robustness against systematic errors compared to the purely dynamic approach. Our method is generalizable to arbitrary two-qubit physical systems, offering a novel and feasible pathway for rapidly and robustly constructing composite two-qubit gates.
		\end{abstract}
		
		\maketitle
		\section{INTRODUCTION}
		In the noisy intermediate-scale quantum (NISQ) era, achieving fast and high-fidelity quantum gates operations is essential for universal quantum computing. In the pursuit of viable quantum computer hardware, spin qubits encoded in silicon double quantum dots (DQDs) are highly promising candidates \cite{hx1,hx2,hx3,hx4,hx5} due to its long coherence time \cite{hx6,hx7}, potential scalability \cite{hx8,hx9} and combination with the modern semiconductor technology \cite{hx10,hx11}. The primary noise sources of silicon-based spin qubits include nuclear and charge noise \cite{hx12,hx13}. Experimentally, the influence of nuclear noise is effectively mitigated benefit from purifying isotope-enriched $^{28}$Si \cite{hx14,hx15,hx16}. However, charge noise remains a critical challenge, as it causes fluctuations in exchange interactions and resonance frequencies, thereby degrading the performance of quantum gates \cite{hx17,hx18}. Nowadays, the gate fidelity on silicon DQDs still needs to be improved to meet the stringent requirements of fault-tolerant quantum computing \cite{hx19,hx20,hx21}.\\
		\indent When building quantum computing circuits, although the universal gate set can be constructed from any single-qubit gate and a nontrivial two-qubit gate \cite{hx22}, many practical algorithms necessitate the use of various two-qubit gates \cite{hx23}, replacing arbitrary two-qubit gates in an algorithm typically requires six to eight single-qubit and three CNOT gates \cite{hx24,hx25,hx26}. Therefore, implementing quantum algorithms through direct sequences of composite two-qubit gates can significantly reduce circuit depth \cite{hx27,hx28,hx29}. It is noteworthy that several useful two-qubit gates operating between two energy levels, inculuding the CZ, CNOT, and iSWAP gates, have already been realized based on the silicon DQDs \cite{hx30,hx31,hx32,hx33,hx34}. When considering composite two-qubit gates involving coherent evolution across more than two energy levels, Ni \emph{et al}. recently experimentally implemented an fSim gate in silicon spin qubits using multi-segment composite pulse sequences \cite{hx35}. In practice, unavoidable environmental noise, control errors, and finite gate duration inherently degrade the accuracy and efficiency of quantum state control. Robust gate construction protocols are therefore urgently required to meet the demands of fault-tolerant quantum computing. Moreover, it is necessary to further enrich the diversity of composite gates that can be realized in DQDs to meet the demands of diverse quantum algorithms.\\
		\indent Among the widely used two-qubit gates, the fSim gate has attracted considerable attention due to its demonstrated utility in various NISQ algorithms. These include quantum approximate optimization algorithms \cite{hx36}, linear depth line algorithms simulating the electronic structure of a molecule \cite{hx37}, and error mitigation techniques \cite{hx38}. Conventionally, the fSim gate is constructed by combining an iSWAP-like gate with a CPHASE gate \cite{hx27,hx39,hx40}. However, this method increases both gate operation time and susceptibility to decoherence, potentially leading to information loss, system instability, and resource inefficiency \cite{hx27}. To ensure that quantum algorithms can be executed within the coherence time of the system, a direct one-step implementation of a high-fidelity fSim gate is highly desirable. In addition, J. Zhan \emph{et al}. proposed a two-qubit B gate in Refs.\cite{hx41}, defined as B=$e^{(i/2)(\pi/2)\sigma_x^1\sigma_x^2}e^{(i/2)(\pi/4)\sigma_y^1\sigma_y^2}$, which enables the implementation of any two-qubit operation with a minimal number of two-qubit and single-qubit gates. As illustrated in Appendix \ref{Appendix A}, a circuit utilizing the B gate can realize a generalized nonlocal two-qubit operation using only two applications of the B gate and six single-qubit gates. This approach offers greater efficiency compared to standard implementations based on CNOT and dual CNOT gates \cite{hx41,hx42,hx43}.\\
		\indent In recent years, various strategies have been proposed to mitigate gate errors, among which optimal control theory \cite{hx44} and geometric quantum computation method \cite{hx45} are prominent examples. Optimal control theory aims to design a set of control pulses that minimize a cost function, with the primary goal often being to maximize gate operation fidelity \cite{hx46}. Additional constraints tailored to specific experimental systems may also be incorporated, such as smoothing the control waveforms \cite{hx47}, restricting pulse amplitudes\cite{hx48}, and accounting for the finite time resolution of arbitrary waveform generators \cite{hx49}. On the other hand, geometric quantum gates utilize geometric phases which depend only on the global evolution path in parameter space to realize robust quantum operations \cite{hx50,hx51,hx52,hx53,hx54,hx55,hx56,hx57}. This global characteristic may offer inherent resilience against certain types of errors, such as those caused by parameter fluctuations in the system Hamiltonian.\\
		\indent In this work, we propose an efficient method for constructing robust two-qubit gates with high-fidelities based on Hamiltonian inverse engineering in silicon DQDs systems. By extending the technique of Hamiltonian inverse engineering with four-dimensional (4D) rotations applied to single electron spin four level systems \cite{hx58,hx59} to two-qubit systems, it enables simultaneous control of state transitions across four energy levels. This capability offers distinct advantages for implementing composite two-qubit gates involving multi-state evolution:  (1) Various composite two-qubit gates are constructed, including a one-step fast-implementable fSim gate scheme and a B gate scheme that only requires a single simple pulse switch. (2) To mitigate the gate fidelity loss under the approximation errors arising from the rotating wave approximation (RWA), we obtain an optimized fSim gate scheme with a continuous and smooth pulse shape by integrating quantum optimal control theory — its average gate time is 50 nanoseconds and theoretical fidelity reaches 99.95\% in the presence of decoherence.  For control errors\cite{hx60}, we propose a geometric+dynamic fSim gate scheme by integrating geometric quantum gate theory. (3) Numerical simulation results show that compared with the pure dynamic implementation scheme, this hybrid gate scheme exhibits significantly enhanced robustness against both types of the Rabi and detuning errors. These findings provide a novel and feasible approach for the fast and high-fidelity implementation of composite two-qubit gates in DQDs systems.\\
		\indent The paper is organized as follows. In Sec. \ref{sec2}, we introduce the Hamiltonian  inverse engineering method based on 4D rotations. In Secs. \ref{sec3}, we present a set of parameterized two-qubit gates based on silicon DQDs, including fSim gate and B gate schemes. In Sec. \ref{sec4}, we analyze the stability of the fSim gate scheme under control errors. In Sec. \ref{sec5}, we construct a geometric+dynamical fSim gate and discuss its robustness. In Sec.\ref{sec6}, we summarize the work of this paper and indicate future prospects for further research.
		\section{Hamiltonian inverse engineering method based on 4D-rotation\label{sec2}}
		For an arbitrary four-level quantum system, the state vector describing the probability information of all observable quantities in the system can be written as
		\begin{align}\label{Eq.1}
			|\psi(t)\rangle=\sum_{n=1}^{n=4}c_n(t)e^{i\varphi_n(t)}|n\rangle,
		\end{align}
		here $c_n$ and $\varphi_n$ represent the amplitude and phase of the n-th energy level, respectively (where we set $\varphi_1=0$). By separating the amplitude and phase components, the state vector can be expressed as $|\psi(t)\rangle=K(t)|\psi_r(t)\rangle $, where
		\begin{align}\label{Eq.2}
			|\psi_r(t)\rangle&=\sum_{n=1}^{n=4}c_n(t)|n\rangle,\nonumber\\
			K(t)&=\sum_{n=1}^{n=4}e^{i\varphi_n(t)}|n\rangle\langle n|,
		\end{align}
		where $|\psi_r(t)\rangle$ is a normalized four-dimensional vector representing the amplitude part of the state vector, and $K(t)$ is a four-dimensional diagonal matrix containing phase information. The evolution operators corresponding to $|\psi(t)\rangle$ and $|\psi_r(t)\rangle$ are respectively defined as $U(t)$ and $U_r(t)$, 
		\begin{align}\label{Eq.3} 
			|\psi(t)\rangle&=U(t)|\psi(0)\rangle,\nonumber\\
			|\psi_r(t)\rangle&=U_r(t)|\psi_r(0)\rangle,
		\end{align}
		the initial evolution time is defined as $t=0$, and the relationship between $U(t)$ and $U_r(t)$ is
		\begin{align}\label{Eq.4}
			U(t)=K(t)U_r(t)K^\dagger(0).
		\end{align}
		According to the Schr\"{o}dinger equation, the Hamiltonian $H$ of the system can be expressed as
		\begin{align}\label{Eq.5}
			H(t)&=i\hbar \dot{U}(t)U^\dagger(t)\nonumber\\
			&=i\hbar K(t) \dot{U}_r(t)U^\dagger_r(t)K^\dagger(t)+i\hbar\dot{K}(t)K^\dagger(t).
		\end{align}
		To construct the parameterized Hamiltonian $H(t)$ and the corresponding evolution operator $U(t)$, we decompose $U_r$ into the product of two isoclinic rotation matrices \cite{hx61}
		\begin{align}\label{Eq.6}
			&U_r(t)=\nonumber\\
			&\left[
			\begin{array}{cccc}
				q_w& -q_x&-q_y&-q_z\\
				q_x& q_w& -q_z& q_y\\
				q_y&q_z & q_w&-q_x\\
				q_z& -q_y& q_x&q_w
			\end{array}\right]
			\left[\begin{array}{cccc}
				p_w& -p_x&-p_y&-p_z\\
				p_x& p_w& p_z& -p_y\\
				p_y&-p_z & p_w&p_x\\
				p_z& p_y& -p_x&p_w
			\end{array}\right],
		\end{align}
		where $q$ and $p$ are two unit quaternions, expressed as $q = q_w + q_x\textbf{i} + q_y\textbf{j} + q_z\textbf{k}$ and $p = p_w + p_x\textbf{i} + p_y\textbf{j} + p_z\textbf{k}$. The above decomposition holds for any four-dimensional rotation matrix, indicating that the construction of $U_r$ is general and not restricted to specific cases. Furthermore, $q$ and $p$ can be represented using a set of four-dimensional azimuths $\{\gamma_i,\theta_i,\phi_i\}$ \cite{hx62}
		\begin{align}\label{Eq.7}
			&q_w=\cos\gamma_1, \ \ \ \ \ \ \ \ \ \ \ \ \ \ \ \ \ \ \ \ \ \ q_x=\sin\gamma_1\cos\theta_1, \nonumber\\
			&q_y=\sin\gamma_1\sin\theta_1\cos\phi_1 , \ \ \ q_z=\sin\gamma_1\sin\theta_1\sin\phi_1,\nonumber\\
			&p_w=\cos\gamma_2 , \ \ \ \ \ \ \ \ \ \ \ \ \ \ \ \ \ \ \ \ \ \ p_x=\sin\gamma_2\cos\theta_2,\nonumber\\
			&p_y=\sin\gamma_2\sin\theta_2\cos\phi_2 ,\ \ \  p_z=\sin\gamma_2\sin\theta_2\sin\phi_2,
		\end{align}
		where $0\leq \phi_{1,2}\leq 2\pi$, $0\leq \theta_{1,2},\gamma_{1,2}\leq 2\pi$. Then substitute the results of the four-dimensional azimuth parameterization $U_r$ into Eqs. (\ref{Eq.4}) and (\ref{Eq.5}), the parameterized evolution matrix $U(t)$ and the parameterized Hamiltonian $H(t)$ can be obtained. Due to the complexity and length of these expressions, they are provided in full in Appendix \ref{Appendix B}.
		\section{composite two-qubit gate based on the silicon DQDs\label{sec3}}
		\subsection{Hamiltonian of silicon DQDs}
\begin{figure}
			\begin{minipage}{0.50\textwidth}
				\centering
				\subfigure{\includegraphics[width=7.0cm]{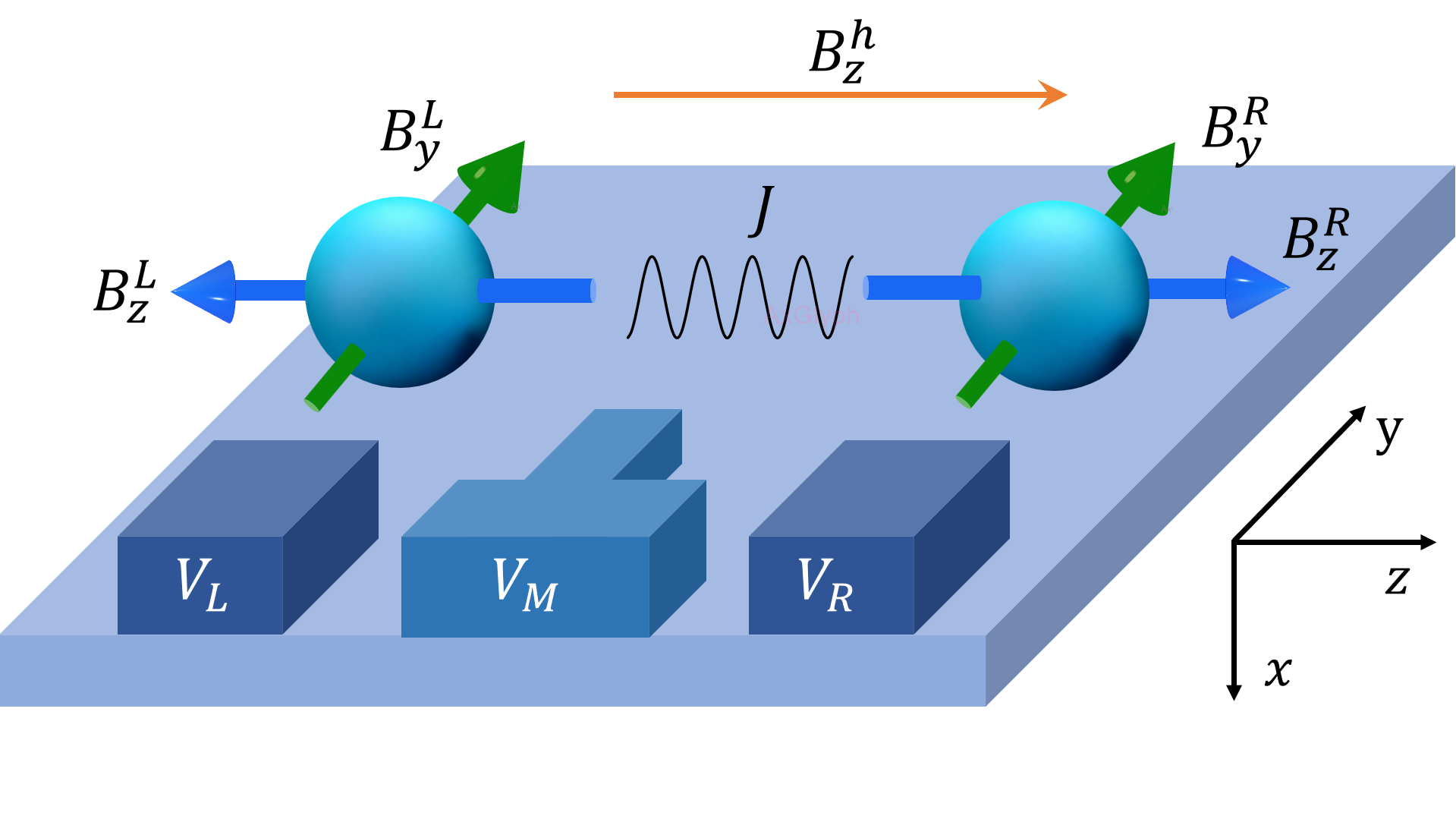}}
			\end{minipage}
			\caption{Spin qubits in the silicon DQDs. The single-qubit gate can be obtained by generating oscillating magnetic fields through the periodic modulation of gate voltages ($V_L$ and $V_R$), while two-qubit gate are realized by precisely regulating the exchange interaction $J$ between spins via the electrostatic barrier gate $V_M$.}
			\label{Y}
		\end{figure}

		The silicon DQDs consists of two quantum dots in a Si/SiGe heterostructure, with one electron confined in each of the left and right quantum dots. The two electrons exhibit an exchange interaction and are experimentally controlled via an applied alternating non-uniform magnetic field, as shown in Fig. \ref{Y}. The corresponding system Hamiltonian is given by
		\begin{align}\label{Eq.8}
		H_D(t)=J(t)(S_L\cdot S_R-1/4)+S_L\cdot B_L+S_R\cdot B_R ,
		\end{align}
		where $J(t)$ denotes the strength of spin exchange between the two electrons, $S_L$ and $S_R$ represent the spin magnitudes of the electrons in the left and right quantum dots, respectively. $B_L=(0,B_y^L(t),B_z^h+B_z^L)^T$ and $B_R=(0,B_y^R(t),B_z^h+B_z^R)^T$ are the magnetic field strengths in the left and right quantum dots, $B_z^h$ is the uniform magnetic field along the z-direction, $B_z^Q(Q=L,R)$ is the non-uniform local magnetic field in the left (right) quantum dots, and $B_y^Q(t) = B_y^{Q,0} + B_y^{Q,1}\cos(\omega t + \varphi) $ represents the transverse oscillating magnetic field. In the two-qubit spin basis $\{|00\rangle , |01\rangle , |10\rangle , |11\rangle\}$, the Hamiltonian in Eq. (\ref{Eq.8}) can be expressed as
		\begin{align}\label{Eq.9}
			H_D(t)=\left[\begin{array}{cccc}
				E_z&-iB_y^R&-iB_y^L&0\\
				iB_y^R&-(\delta E_z+J)/2&J/2&-iB_y^L\\
				iB_y^L&J/2&(\delta E_z-J)/2&-iB_y^R\\
				0&iB_y^L&iB_y^R&-E_z
			\end{array}\right],
		\end{align}
		where $E_z=B_z^h+(B_z^R+B_z^L)/2$ represents the average Zeeman splitting magnitude, and $\delta E_z=B_z^R-B_z^L$ denotes the difference in Zeeman splitting between the two quantum dots. \\
		\indent For subsequent numerical simulations, we adopt the following experimental data of silicon DQDs recently\cite{hx34,hx63}: $E_z/2\pi = 20.64$ GHz, $\delta E_z/2\pi = 214$ MHz, $B_y^{L,0}/2\pi = 5$ MHz, $B_y^{R,0} /2\pi = 55$ MHz, $J/2\pi = 19.7$ MHz. 
		\subsection{One-step implementable parameterized two-qubit quantum gate set}
		The Hamiltonian of the silicon DQDs in Eq. (\ref{Eq.9}) possesses the following symmetry
		\begin{align}\label{Eq.10}
			\Omega_{12}=\Omega_{34},\ \ \ \Omega_{13}=\Omega_{24}, \ \ \ \ \Omega_{14}=0,
		\end{align}
		where $\Omega_{ij}$ represents the amplitude of the component of the Hamilton quantity in row $i$ and column $j$. Substituting it into Eq. (\ref{Eq.B.2}),
		a set of four-dimensional azimuth parameters satisfying the conditions can be selected as
		\begin{align}\label{Eq.11}
			&\dot{\phi_1}=0,\ \ \ \ \dot{\phi_2}=0,\ \ \ \ \dot{\theta_1}=0,\ \ \ \ \dot{\theta_2}=0,\nonumber\\
			&\phi_1=\frac{\pi}{2},\ \ \ \ \theta_2=\frac{\pi}{2}, \ \ \ \ \dot{\gamma}_1(t)\sin\theta_1=-\dot{\gamma}_2(t)\sin\phi_2,
		\end{align}
		with the above parameters substituted, the Hamiltonian of the silicon DQD system can be cast into its parameterized form as follows
		\begin{align}\label{Eq.12}
			H_p(t)&=-\dot{\varphi}_2(t)|01\rangle\langle01|-\dot{\varphi}_3(t)|10\rangle\langle10|-\dot{\varphi}_4(t)|11\rangle\langle11|\nonumber\\
			&-i\dot{\gamma}_1(t)\cos\theta_1e^{-i\varphi_2}|00\rangle\langle01|\nonumber\\
			&+i\dot{\gamma}_1(t)\sin\theta_1\cot\phi_2e^{-i\varphi_3}|00\rangle\langle10|\nonumber\\
			&-i \dot{\gamma}_1(t)\sin\theta_1e^{i(\varphi_2-\varphi_3)}|01\rangle\langle10|\nonumber\\
			&+i\dot{\gamma}_1(t)\sin\theta_1\cot\phi_2e^{i(\varphi_2-\varphi_4)}|01\rangle\langle11|\nonumber\\
			&+i\dot{\gamma}_1(t)\cos\theta_1e^{-i(\varphi_3-\varphi_4)}|10\rangle\langle11|+ \mathrm{H.c.}
		\end{align}
		By substituting the parameter conditions in Eq. (\ref{Eq.11}) into Eq. (\ref{Eq.6}) and Eq. (\ref{Eq.7}), the parameterized total evolution matrix can be constructed as
		\begin{align}\label{Eq.13}
			&U_p(t,0)=K(t)U_rK^\dagger(0),\nonumber\\
			&\quad \quad \quad \ =K(t)M_LM_RK^\dagger(0),\nonumber\\
			&M_L=\nonumber\\
			&\!\!\!\!\!\!\!\!\!\left[\begin{array}{cccc}	
				\cos\gamma_1&-\cos\theta_1\sin\gamma_1&0&-\sin\gamma_1\sin\theta_1\\
				\cos\theta_1\sin\gamma_1&\cos\gamma_1&-\sin\gamma_1\sin\theta_1&0\\
				0&\sin\gamma_1\sin\theta_1&\cos\gamma_1&-\cos\theta_1\sin\gamma_1\\
				\sin\gamma_1\sin\theta_1&0&\cos\theta_1\sin\gamma_1&\cos\gamma_1
			\end{array}\right],\nonumber\\
			&M_R=\nonumber\\
			&\!\!\!\!\!\!\!\!\!\left[\begin{array}{cccc}
				\cos\gamma_2&0&-\cos\phi_2\sin\gamma_2&-\sin \phi_2\sin\gamma_2\\
				0&\cos\gamma_2&\sin\phi_2\sin\gamma_2&-\cos\phi_2\sin\gamma_2\\
				\cos\phi_2\sin\gamma_2&-\sin\phi_2\sin\gamma_2&\cos\gamma_2&0\\
				\sin\phi_2\sin\gamma_2&\cos\phi_2\sin\gamma_2&0&\cos\gamma_2
			\end{array}\right].
		\end{align}
		To ensure the initial condition $U(0,0)=I$, the parameters must satisfy $\gamma_1(0)=\gamma_2(0)=0$, which implies the relation $\gamma_1(t)\sin\theta_1=-\gamma_2(t)\sin\phi_2$. The operator $U_p$ represents the parameterized two-qubit gate set enabling one-step implementation in the silicon DQDs, where $U_r=M_LM_R$ is amplitude part of $U_p$. The $U_r$ possesses three independent degrees of freedom, a feature dictated by the symmetry of the actual Hamiltonian in Eq. (\ref{Eq.9}).
	
		\subsection{fSim gate}
		The fSim gate is represented by the matrix in the basis $\{|00\rangle, |01\rangle, |10\rangle, |11\rangle\}$ as follows:
		\begin{align}\label{Eq.14}
			\mathrm {fSim}(\vartheta,\Xi)=\left[\begin{array}{cccc}
				1&0&0&0\\
				0&\cos\vartheta&-i\sin\vartheta&0\\
				0&-i\sin\vartheta&\cos\vartheta&0\\
				0&0&0&e^{i\Xi}
			\end{array}\right],
		\end{align}
		the standard approach to constructing the fSim gate involves combining an iSWAP-like gate with a CPHASE gate. However, this method typically requires complex pulse switching and leads to inefficient use of gate time. In this section, we present a scheme that enables direct one-step implementation of the fSim gate in silicon DQDs. Our procedure consists of the following steps: first, make parameterized evolution matrix $U_p$ form an fSim gate and solve for the specific 4D azimuthal angle $\{\gamma_{1,2},\theta_{1,2},\phi_{1,2}\}$ and phase $\{\varphi_{1,2,3,4}\}$; then, substitute these parameters to obtain the parameterized Hamiltonian $H_p$; finally, match the actual Hamiltonian $H$ with the $H_p$ and derive the corresponding constraints on the physical parameters.\\
		\indent Setting $U_p(T,0)=\mathrm{fSim}(\vartheta,\Xi)$, where the system evolves from initial time $t=0$ to final time $t=T$. When only considering the amplitude component, the following condition is satisfied: $U_{r[1,2]}=U_{r[1,3]}=U_{r[1,4]}=0$, where $U_{r[i,j]}$ represents the row $i$ and column $j$ component of matrix $U_r$. Substituting these quantities into Eq. (\ref{Eq.13}), one can obtain
		\begin{align}\label{Eq.15}
			\phi_2=\pi/2, \ \ \ \ \theta_1=\pi/2,\ \ \ \ \gamma_1(t)=-\gamma_2(t). 
		\end{align}
		Consequently, the expression for $U_p$ is simplified to
		\begin{align}\label{Eq.16}
			U^f_p(T,0)=
			\left[\begin{array}{cccc}
				1&0&0&0\\
				0&\cos \gamma_1 e^{i\varphi_{22}}&-\sin\gamma_1 e^{i\varphi_{23}}&0\\
				0&\sin\gamma_1 e^{i\varphi_{32}}&\cos\gamma_1 e^{i\varphi_{33}}&0\\
				0&0&0&e^{i\varphi_{44}}
			\end{array}\right],
		\end{align}
		where $\varphi_{ij}=\varphi_i(T)-\varphi_j(0)$, and it requires that $U(0,0)=I$, i.e $\gamma_1(0)=0$. Equating the expressions in Eq. (\ref{Eq.14}) and Eq. (\ref{Eq.16}) allows the phase parameters to be determined as follows
		\begin{align}\label{Eq.17}
			&\varphi_2(T)-\varphi_2(0)=2n_1\pi,\quad \varphi_3(T)-\varphi_3(0)=2n_2\pi,\nonumber\\
			&\varphi_4(T)-\varphi_4(0)=\Xi+2n_3\pi,\nonumber\\
			&\varphi_2(T)-\varphi_3(0)=\pi/2+2n_4\pi,\nonumber\\
			&\varphi_3(T)-\varphi_2(0)=-\pi/2+2n_5\pi,\quad \gamma_1(T)=\vartheta,
		\end{align}
		where $n_i\in Z$. The parameterized Hamiltonian $H_p$ is expressed as 
		\begin{align}\label{Eq.18}
			H^f_p(t)=\left[\begin{array}{cccc}
				0& 0&0&0\\
				0&-\dot{\varphi}_2&-i\dot{\gamma}_1e^{i(\varphi_2-\varphi_3)}&0\\
				0&i\dot{\gamma}_1e^{i(\varphi_3-\varphi_2)} & -\dot{\varphi}_3&0\\
				0&0&0&-\dot{\varphi}_4
			\end{array}\right].
		\end{align}
		\indent In order to match the actual Hamiltonian $H_D(t)$ of the silicon DQDs with the above $H^f_p$, the actual Hamiltonian needs to be reconstructed. By setting $B_y^R(t)=B_y^L(t)=0,J=2j\cos(\omega t+\psi)$, Eq. (\ref{Eq.9}) can be written as
		\begin{widetext}
			\begin{align}\label{Eq.19}
				H_f(t)=\left[\begin{array}{cccc}
					E_z&0&0&0\\
					0&-\delta E_z/2-j\cos(\omega t+\psi)&j\cos(\omega t+\psi)&0\\
					0&j\cos(\omega t+\psi)&\delta E_z/2-j\cos(\omega t+\psi)&0\\
					0&0&0&-E_z
				\end{array}\right].
			\end{align}
			\vspace{-0.5em} 
		\end{widetext}
		A rotation frame transformation applied to this Hamiltonian gives 
		\begin{align}\label{Eq.20}
			H'_f(t)&=U_t^\dagger(t)H_f(t)U_t(t)-i\hbar U_t^\dagger\frac{d}{dt}U_t(t),\nonumber\\
			&=\left[\begin{array}{cccc}
				E_z&0&0&0\\
				0&-j\cos(\omega t)&j(1+e^{-i2wt})/2&0\\
				0&j(1+e^{i2wt})/2&-j\cos(\omega t)&0\\
				0&0&0&-E_z
			\end{array}\right],
		\end{align}
		where  transformation matrix $U_t=\mathrm{exp}[-i(\omega t/4)I\otimes\sigma^2_z+i(\omega t/4)\sigma^1_z\otimes I]$, and set $\psi=0,\omega=\delta E_z$. Hereafter and in the remainder of this manuscript, we set $\hbar=1$. Applying the RWA to discard the high-frequency oscillation term ($j(t)e^{\pm i2wt}$) , and setting the ground state energy to zero, the Eq. (\ref{Eq.20}) can be described as
		\begin{align}\label{Eq.21}
			H''_f(t)=\left[\begin{array}{cccc}
				0&0&0&0\\
				0&-j\cos(\omega t)-E_z&j/2&0\\
				0&j/2&-j\cos(\omega t)-E_z&0\\
				0&0&0&-2E_z
			\end{array}\right].
		\end{align}
		Comparing the actual Hamiltonian in Eq. (\ref{Eq.21}) with the parameterized Hamiltonian in Eq. (\ref{Eq.18}) gives rise to
		\begin{align}\label{Eq.22}
			&\dot{\varphi_2}(t) =\dot{\varphi_3}(t)= E_z+j\cos(\omega t), \nonumber\\
			&\dot{ \varphi_4}(t) = 2E_z,\quad
			-i\dot{\gamma}_1e^{i(\varphi_2(t)-\varphi_3(t))}=\frac{j(t)}{2},
		\end{align}
		\begin{figure}
			\begin{minipage}{0.23\textwidth}
				\centering
				\includegraphics[width=4.5cm]{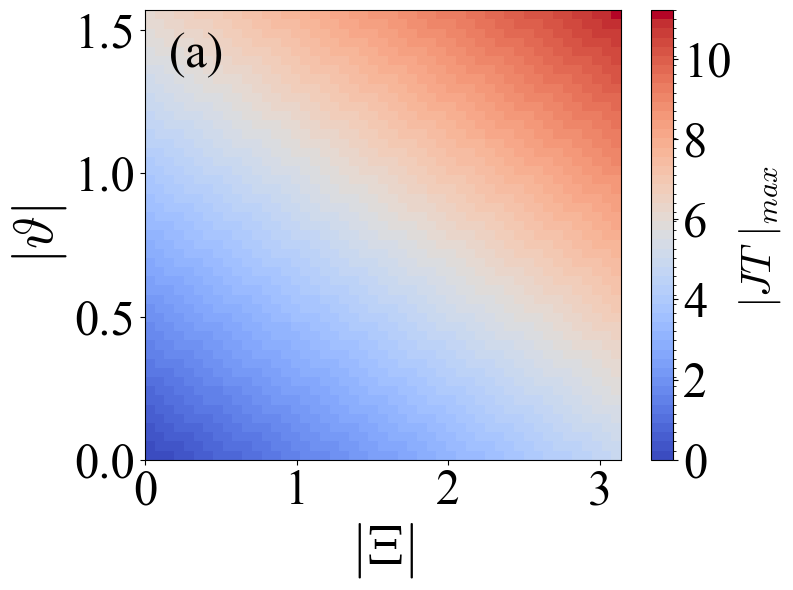} 
			\end{minipage}
			\hfill 
			\begin{minipage}{0.23\textwidth}
				\centering
				\includegraphics[width=4cm]{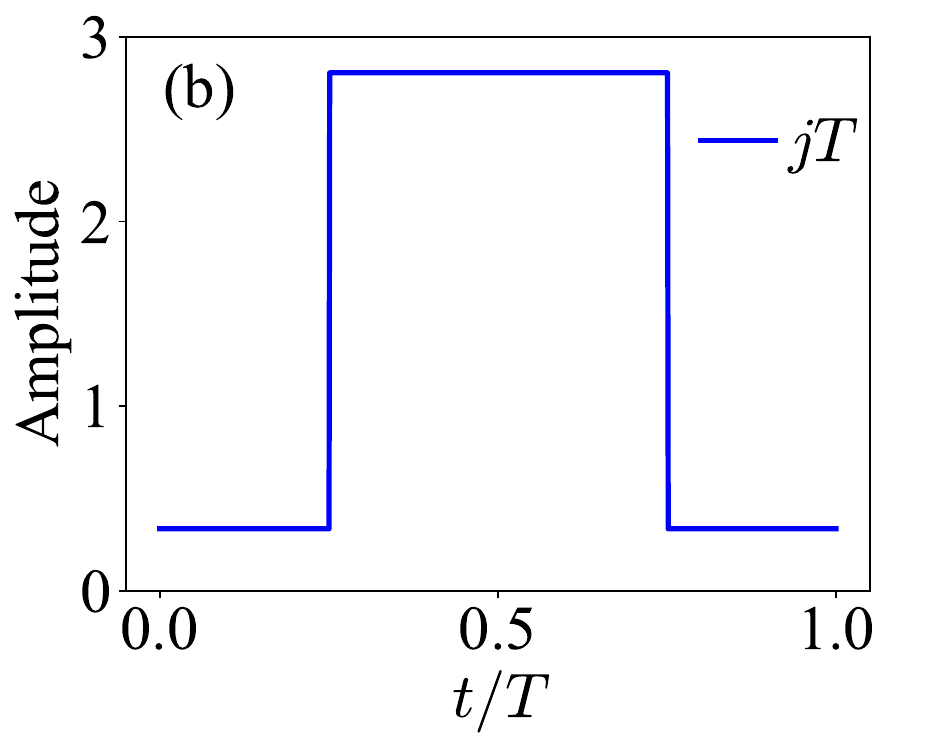} 
			\end{minipage}
			\caption{(a) The maximum value of pulse amplitude $|JT|_{max}$ with different fSim gate parameters $\{\vartheta, \Xi\}$ under the rectangular pulse scheme. Where the parameter ranges selected are $|\vartheta|\in[0,\pi/2]$ and $|\Xi|\in[0,\pi]$. (b) The $jT$ under the rectangular pulse scheme of fSim gate, where fSim gate parameters are $\vartheta=\pi/4,\Xi=\pi/2$, gate time $T\approx45$ ns, $\delta E_z \approx 2\pi\times 22$ MHz and $E_z\approx 2\pi\times 5.6$ MHz.}
			\label{f1}
		\end{figure}
		substituting these results into Eq. (\ref{Eq.17}) allows us to determine the actual physical parameters necessary for implementing the fSim gate as follows
		\begin{align}\label{Eq.23}
			&E_z=\frac{\Xi}{2T},\quad \delta E_z=\omega ,\quad J=2j\cos(\omega t)\nonumber\\
			&\int_{0}^{T}j(t)\cos(\omega t)dt=-\frac{\Xi}{2},\quad\int_{0}^{T}j(t)dt=2\vartheta.
		\end{align}
		\indent Note that no constraints have been imposed on the value of $\delta E_z$ in this formulation.  To attempt to minimize the approximation error and simplify the calculation, we set $\delta E_z=2N\pi/T$, where $N$ is an integer. Once $\delta E_z$ is determined, the magnitude of $JT$ becomes proportional to the fSim gate parameter $\{\vartheta,\Xi\}$. To shorten the gate duration $T$, we choose $\vartheta \in [-\pi/2, \pi/2]$ and $\Xi \in [-\pi, \pi]$. This range is complete for $\Xi$, while for $\vartheta$, it suffices to set $E_z = \Xi/(2T) + \pi/T$. Under these choices, the resulting evolution matrix takes the form
		\begin{align}\label{Eq.24}
			U'(\vartheta,\Xi)&=\left[\begin{array}{cccc}
				1&0&0&0\\
				0&-\cos\vartheta&i\sin\vartheta&0\\
				0&i\sin\vartheta&-\cos\vartheta&0\\
				0&0&0&e^{i\Xi}
			\end{array}\right]\nonumber\\
			&=\mathrm{fSim}(\vartheta+\pi,\Xi),
		\end{align}
		this constitutes a complete fSim gate set. We further present a specific gate construction scheme using a simple rectangular $j(t)$ pulse
		\begin{align}\label{Eq.25}
			&j=(8\vartheta-\Xi\pi)/(4T),\quad 0\le t<T/4,\nonumber\\
			&j=(8\vartheta+\Xi\pi)/(4T),\quad T/4\le t< 3T/4,\nonumber\\
			&j=(8\vartheta-\Xi\pi)/(4T),\quad 3T/4\le t \le T,\nonumber\\
			&J(t)=2j(t)\cos(\delta E_z t),
		\end{align}
		where $\delta E_z$ is chosen as $\delta E_z=2\pi/T$ and $E_z=\Xi/(2T)$. Substituting the experimental data shows that the gate time is primarily constrained by the magnitude of the exchange interaction $J$. Fig. \ref{f1}(a) illustrates the maximum pulse amplitude $|JT|_{max}$ for different fSim gate parameters $\{\vartheta, \Xi\}$. It can be observed from Fig. \ref{f1}(a) that the fSim gate time under the rectangular scheme is proportional to the gate parameters. Using $J=2\pi\times 19.7$ MHz, we find that the fSim gate time under the rectangular pulse scheme does not exceed 90 ns for any parameter set, with an average calculated gate time of 45 ns. We also present the shape of a specific pulse $jT$ as shown in Fig. \ref{f1}(b), where the fSim gate parameters are chosen as $(\vartheta=\pi/4,\Xi=\pi/2)$, equivalent to combining a $\sqrt{\mathrm{iSWAP}}$ gate and a $\sqrt{\mathrm{CZ}}$ gate.
		\subsection{B gate}
		The B gate enables the implementation of any two-qubit operation using a minimal number of both two-qubit and single-qubit gates. Its matrix representation in the $\{|00\rangle, |01\rangle, |10\rangle, |11\rangle\}$ basis is given by
		
		\begin{align}\label{Eq.26}
			B=\left[\begin{array}{cccc}
				\cos(\frac{\pi}{8})&0&0&i\sin(\frac{\pi}{8})\\
				0&\sin(\frac{\pi}{8})&i\cos(\frac{\pi}{8})&0\\
				0&i\cos(\frac{\pi}{8})&\sin(\frac{\pi}{8})&0\\
				i\sin(\frac{\pi}{8})&0&0&\cos(\frac{\pi}{8})
			\end{array}\right].
		\end{align}

\begin{figure}
		\begin{minipage}{0.45\textwidth}
			\centering
			\subfigure{\includegraphics[width=8cm]{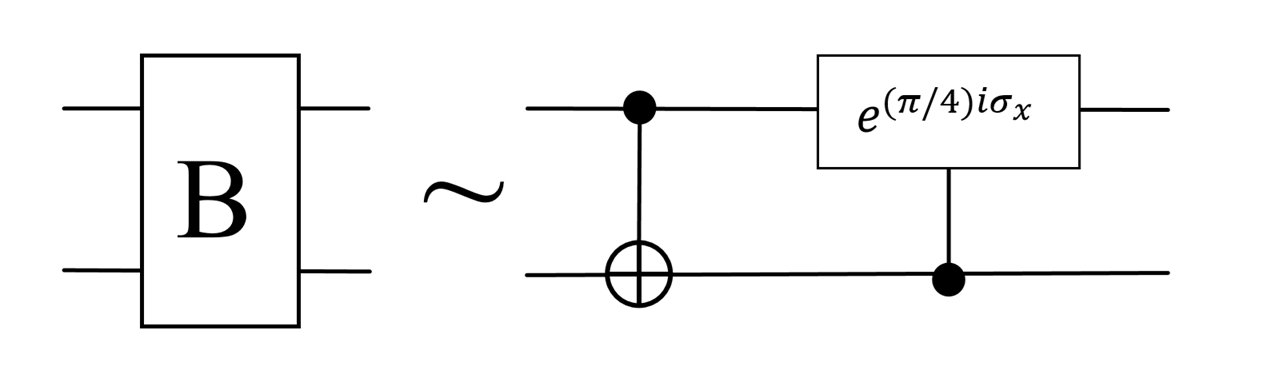}} 
		\end{minipage}
		\caption{A general construction scheme for B gate, combining a CNOT gate and a controlled-$e^{(\pi/4)i\sigma_x}$ gate.}
		\label{f2}
	\end{figure}

\begin{figure}
			\begin{minipage}{0.24\textwidth}
				\centering
				\subfigure{\includegraphics[width=4.3cm]{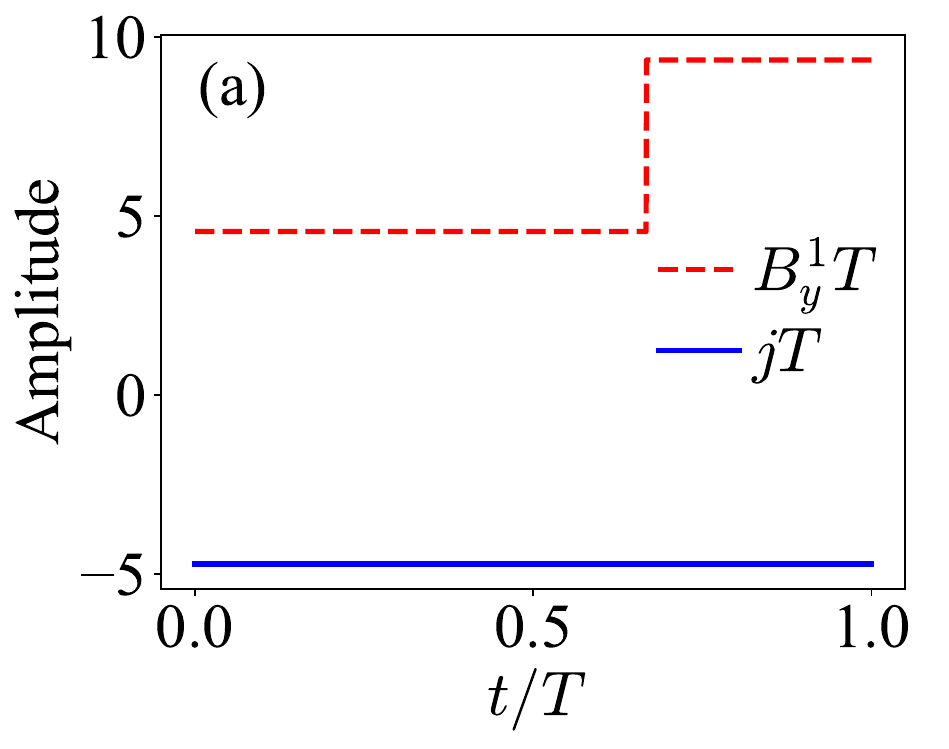}} 
			\end{minipage}\begin{minipage}{0.24\textwidth}
				\centering
				\subfigure{\includegraphics[width=4.3cm]{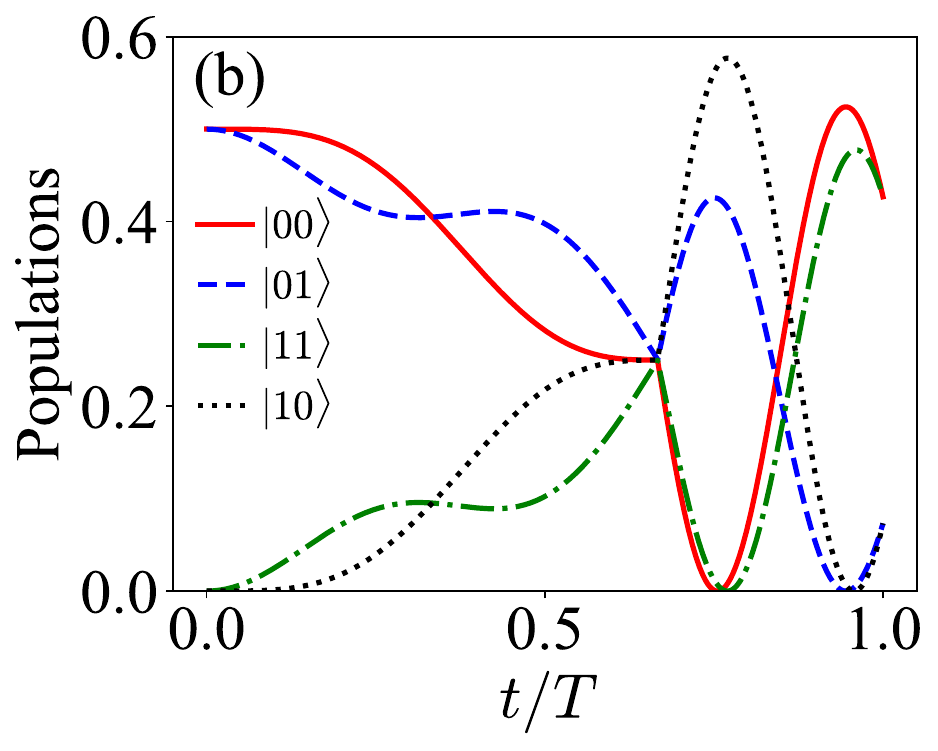}} 
			\end{minipage}
			\caption{(a) The $jT, B^1_yT$ under the rectangular pulse scheme of B gate. (b) Population transfer during the B gate operation. The initial state is chosen as $(|00\rangle+|01\rangle)/\sqrt{2}$, where the horizontal axes of both subfigures represent the gate operation duration, and the gate time $T\approx 76$ ns.}
			\label{f3}
		\end{figure}

The general approach to constructing the B gate involves combining a CNOT gate with a controlled-$e^{(\pi/4)i\sigma_x}$ gate \cite{hx41}, as illustrated in Fig. \ref{f2}. However, since the control qubits for these two gates differ, their simultaneous implementation within the same system typically requires complicated pulse switching, thereby increasing experimental overhead. In this section, we propose a B gate implementation scheme tailored to silicon DQDs, which can be realized using only a single simple pulse switch.\\ 
\indent The B gate can be constructed by combining a $B_1$ gate and a $B_2$ gate, where $B_1(\varGamma_1)=-e^{i\varGamma_1 \sigma^1_x\sigma^2_x}$ and $
B_2(\varGamma_2)=-e^{i\varGamma_2\sigma^1_y\sigma^2_y }$ with $\varGamma_1=\pi/4$ and $\varGamma_2=\pi/8$. The matrix representations in the computational basis are given as follows\\	
		\begin{align}\label{Eq.27}
			B_1(\varGamma_1)&=-\left[\begin{array}{cccc}
				\cos\varGamma_1&0&0&i\sin\varGamma_1\\
				0&\cos\varGamma_1&i\sin\varGamma_1&0\\
				0&i\sin\varGamma_1&\cos\varGamma_1&0\\
				i\sin\varGamma_1&0&0&\cos\varGamma_1
			\end{array}\right],\nonumber\\
			B_2(\varGamma_2)&=-\left[\begin{array}{cccc}
				\cos\varGamma_2&0&0&i\sin\varGamma_2\\
				0&\cos\varGamma_2&-i\sin\varGamma_2&0\\
				0&-i\sin\varGamma_2&\cos\varGamma_2&0\\
				i\sin\varGamma_2&0&0&\cos\varGamma_2
			\end{array}\right].
		\end{align}	
	Setting $U_p(T,0)=B_i(\varGamma_i)(i=1,2)$, where the system evolves from initial time $t=0$ to final time $t=T$. When only the amplitude part is considered, the following constraints on the matrix elements must be satisfied: $U_{r[1,2]}=U_{r[1,3]}=0, U_{r[1,4]}=U_{r[2,3]}$, where $U_{r[i,j]}$ represents the in row $i$ and column $j$ component of $U_r$,  and $U_r$ represents the amplitude part of the total evolution matrix. Substituting the above results into Eq. (\ref{Eq.13}) allows us to solve for the corresponding parameters
		\begin{align}\label{Eq.28}
			\gamma_1(T)=\pi, \quad \phi_2=\frac{\pi}{2}, \quad \gamma_2(t)=-\sin\theta_1\gamma_1(t).
		\end{align}
	Therefore, the $U_p$ reads
	\begin{align}\label{Eq.29}
			&U^B_p(T,0)=\nonumber\\
			&-\left[\begin{array}{cccc}
				\cos\gamma_2&0&0&\sin\gamma_2 e^{-i\varphi_{14}}\\
				0&\cos\gamma_2 e^{i\varphi_{22}}&\sin\gamma_2 e^{i\varphi_{23}}&0\\
				0&-\sin\gamma_2 e^{i\varphi_{32}}&\cos\gamma_2 e^{i\varphi_{33}}&0\\
				\sin\gamma_2 e^{i\varphi_{41}}&0&0&\cos\gamma_2 e^{i\varphi_{44}}
			\end{array}\right],
		\end{align}
		where $\varphi_{ij}=\varphi_i(T)-\varphi_j(0),\varphi_1(t)=0$, and the initial condition $U(0,0)=I$, i.e. $\gamma_2(0)=0$. To construct a $B_1(\varGamma_1)$ gate, from Eqs. (\ref{Eq.27}) and (\ref{Eq.29}), we can get the following solutions as
		\begin{align}\label{Eq.30}
			&\varphi_2(0)=\pi/2,\quad \varphi_2(T)=\pi/2+2n_1\pi,\nonumber\\
			&\varphi_3(0)=0, \quad \ \ \ \ \varphi_3(T)=2n_2\pi,\nonumber\\ 
			&\varphi_4(0)=\pi/2, \quad \varphi_4(T)=\pi/2+2n_3\pi, \nonumber\\
			&\gamma_2(T)=\varGamma_1,\quad \ \ \ \sin\theta_1=-\varGamma_1/\pi,
		\end{align}
		where $n_i\in Z$. To construct the $B_2(\vartheta_2)$ gate, it is sufficient to take $\varphi(0)=-\pi/2, \varphi_4(T)=-\pi/2+2n_3\pi$ and replace $\varGamma_1$ by $\varGamma_2$. Under these conditions, the parameterized Hamiltonian $H_p$ is reduced to the following form
		\begin{widetext}
			\begin{align}\label{Eq.31}
			H^B_p(t)=\left[\begin{array}{cccc}
					0& i\dot{\gamma}_2\cot\theta_1e^{-i\varphi_2(t)}&0&0\\
					-i\dot{\gamma}_2\cot\theta_1e^{i\varphi_2(t)}&-\dot{\varphi}_2&i 2\dot{\gamma}_2e^{i(\varphi_2(t)-\varphi_3(t))}&0\\
					0&-i2\dot{\gamma}_2e^{-i(\varphi_2(t)-\varphi_3(t))} & -\dot{\varphi}_3&i\dot{\gamma}_2\cot\theta_1e^{i(\varphi_3(t)-\varphi_4(t))}\\
					0&0&-i\dot{\gamma}_2\cot\theta_1e^{-i(\varphi_3(t)-\varphi_4(t))}&-\dot{\varphi}_4
				\end{array}\right].
			\end{align}
			\vspace{-0.5em} 
		\end{widetext}
		To align the physical Hamiltonian in Eq. (\ref{Eq.9}) with its parameterized form in Eq. (\ref{Eq.31}), some extra constraints must be imposed on the actual Hamiltonian. We focus on the case of weak exchange, i.e. $J\ll \delta E_z$ \cite{hx33,hx34,hx63}, and set $B_y^L(t)=0$, the actual Hamiltonian becomes \cite{hx30}
		\begin{align}\label{Eq.32}
			H_B=\left[\begin{array}{cccc}
				E_z&-iB_y^R(t)&0&0\\
				iB_y^R(t)&-\delta E_z/2&J(t)/2&0\\
				0&J(t)/2&\delta E_z/2&-iB_y^R(t)\\
				0&0&iB_y^R(t)&-E_z
			\end{array}\right],
		\end{align}
		then set $B_y^R(t)=B_y^{1}(t)\cos(\omega_2t+\psi_1),J=2j(t)\cos[(\omega_1-\omega_2)t]$, and perform a rotating frame transformation and the RWA on the Hamiltonian in Eq. (\ref{Eq.32}). The transformation matrix is $U'_t=\mathrm{exp}[-i(\omega_1 t/2)\sigma^1_z\otimes I-i(\omega_2 t/2)I\otimes\sigma^2_z]$. The transformed Hamiltonian is written in the new $\{|\widetilde{00}\rangle, |\widetilde{01}\rangle, |\widetilde{10}\rangle, |\widetilde{11}\rangle\}$ basis as
		\begin{align}\label{Eq.33}
			H'_B=\left[\begin{array}{cccc}
				0&-iB_y^{1}e^{i\psi_1}&0&0\\
				iB_y^{1}e^{-i\psi_1}&0&j/2&0\\
				0&j/2&0&-iB_y^{1}e^{i\psi_1}\\
				0&0&iB_y^{1}e^{-i\psi_1}&0
			\end{array}\right],
		\end{align}
		where the parameters set as $\omega_1+\omega_2=2E_z,\omega_1-\omega_2=-\delta E_z$. By comparing the actual Hamiltonian in Eq. (\ref{Eq.33}) with the parameterized Hamiltonian in Eq. (\ref{Eq.31}), and substituting these results into Eq. (\ref{Eq.30}), a set of physical parameters for constructing a $B_1(\varGamma_1)$ gate are solved as
		\begin{align}\label{Eq.34}
			&\int_{0}^{T}j(t)dt=-4\varGamma_1, \quad B_y^1=j\cot(\arcsin(-\varGamma_1/\pi))/4,   \nonumber\\
			&J(t)= 2j(t)\cos[(\omega_1-\omega_2)t],\quad E_z=(\omega_1+\omega_2)/2,\nonumber\\
			&B_y^R(t)=B_y^1\cos(\omega_2t+\pi/2),\quad \delta E_z=\omega_2-\omega_1.
		\end{align}
		Similarly, a set of physical parameters for constructing a $B_2(\varGamma_2)$ gate are 
		\begin{align}\label{Eq.35}
			&\int_{0}^{T}j(t)dt=-4\varGamma_2, \quad B_y^1=j\cot(\arcsin(-\varGamma_2/\pi))/4,   \nonumber\\
			&J(t)= 2j(t)\cos[(\omega_1-\omega_2)t],\quad E_z=(\omega_1+\omega_2)/2,\nonumber\\
			&B_y^R(t)=B_y^1\cos(\omega_2t),\quad \delta E_z=\omega_2-\omega_1.
		\end{align}
		When we choose $\varGamma_1=\pi/4$ and $\varGamma_2=\pi/8$, a complete B gate is constructed by combining a $B_1$ gate and a $B_2$ gate. Here, a specific B gate construction scheme with simple rectangular pulses are designed as
		\begin{align}\label{Eq.36}
			&B_y^R(t)=B_y^1\cos[(E_z+\delta E_z/2)t+\pi/2], \nonumber\\
			&B_y^1=\frac{-3\pi\cot(\arcsin(-1/4))}{8T},\quad 0\le t <\frac{2T}{3}\nonumber\\
			&B_y^R(t)=B_y^1\cos[(E_z+\delta E_z/2)t],\nonumber\\
			&B_y^1=\frac{-3\pi\cot(\arcsin(-1/8))}{8T},\quad \frac{2T}{3}\le t \le T,\nonumber\\
			&j=\frac{-3\pi}{2T},\quad J(t)= 2j\cos(\delta E_z t).
		\end{align}
		Notably, no constraints are imposed on the values of $E_z,\delta E_z$, and larger values of these parameters are generally beneficial for satisfying the approximation conditions. The temporal profiles of the pulses $j(t)$ and $B_y^1(t)$ are shown in Fig. \ref{f3}(a) and the corresponding population transfer dynamics are displayed in Fig. \ref{f3}(b). The evolution matrix in the interval $0\le t<2T/3$ constitutes a $B_1(\frac{\pi}{4})$ gate, while that in $2T/3\le t\le T$ realizes a $B_2(\frac{\pi}{8})$ gate. Only a single switch of the magnetic field $B_y^R$ is required between the two evolution processes. Using the experimental parameters, the total gate time for the B gate is found to be $T\approx 76$ ns. For comparison, the CNOT gate under the same experimental conditions requires 94 ns in silicon DQDs \cite{hx33}. Therefore, the B gate offers a more efficient pathway for constructing quantum circuits, enabling implementations with fewer quantum gates and shorter overall execution times.
	   \section{gate performance\label{sec4}}
	   \subsection{Approximation error}
\begin{figure}
	   \begin{minipage}{0.23\textwidth}
	   	\centering
	   	\subfigure{\includegraphics[width=4.3cm]{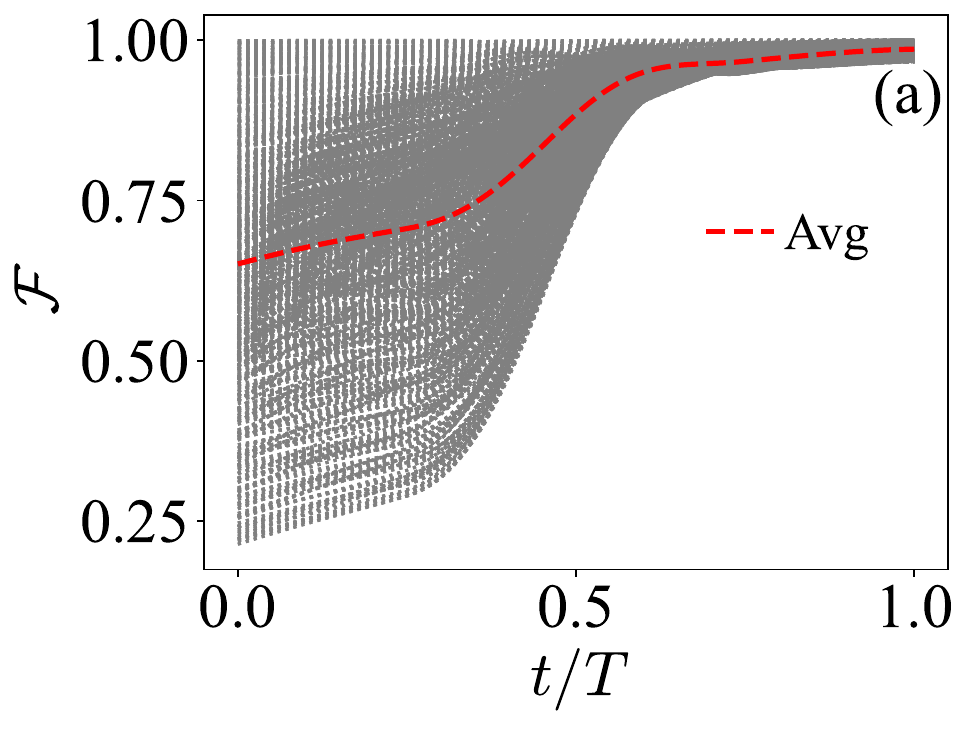}} 
	   \end{minipage}
	   \hfill 
	   \begin{minipage}{0.24\textwidth}
	   	\centering
	   	\subfigure{\includegraphics[width=4cm]{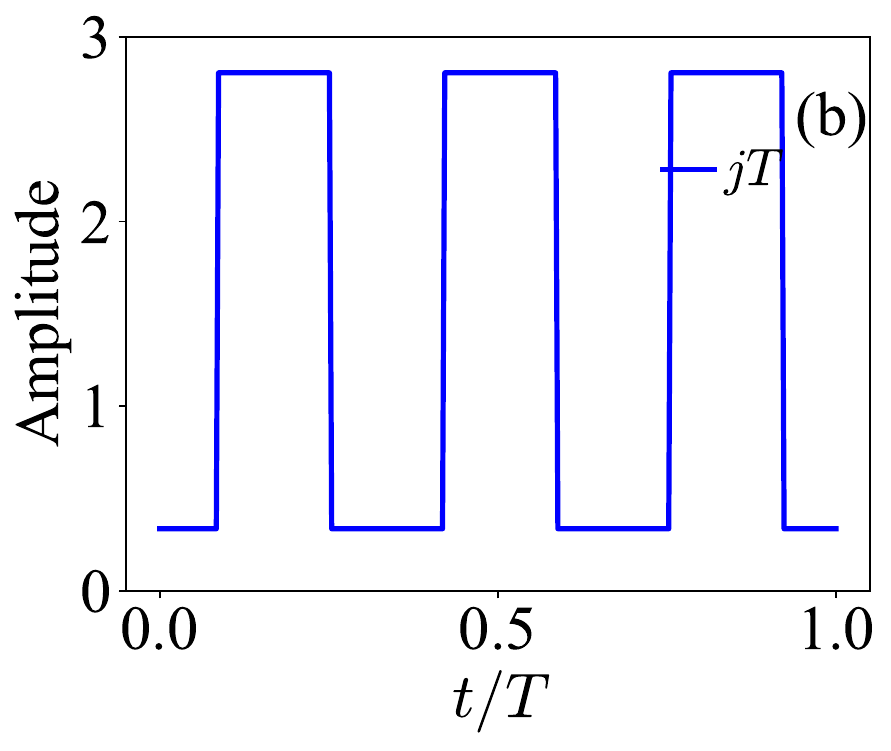}} 
	   \end{minipage}
    \begin{minipage}{0.24\textwidth}
   	\centering
   	\subfigure{\includegraphics[width=4cm]{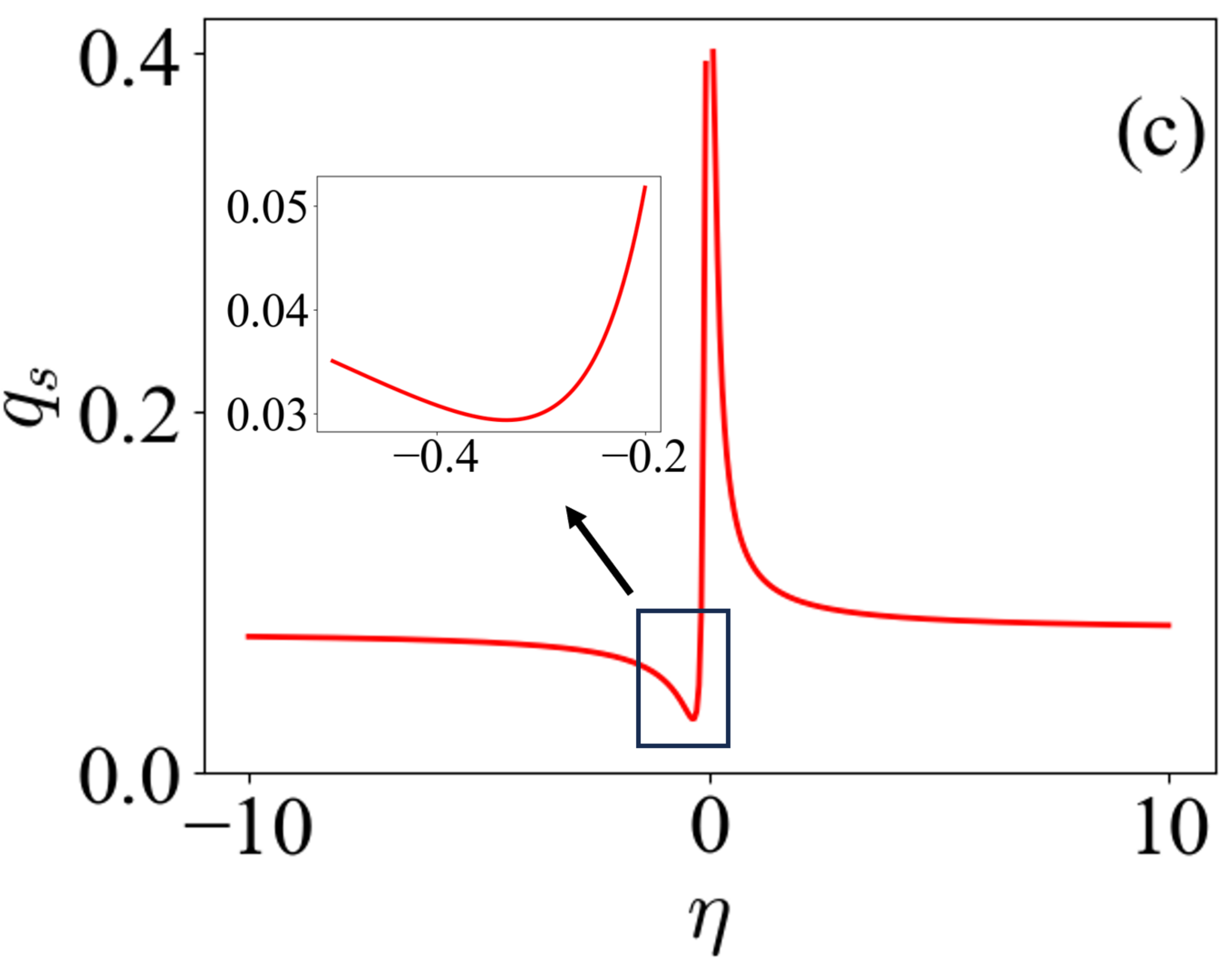}} 
   \end{minipage}\begin{minipage}{0.24\textwidth}
   	\centering
   	\subfigure{\includegraphics[width=4.3cm]{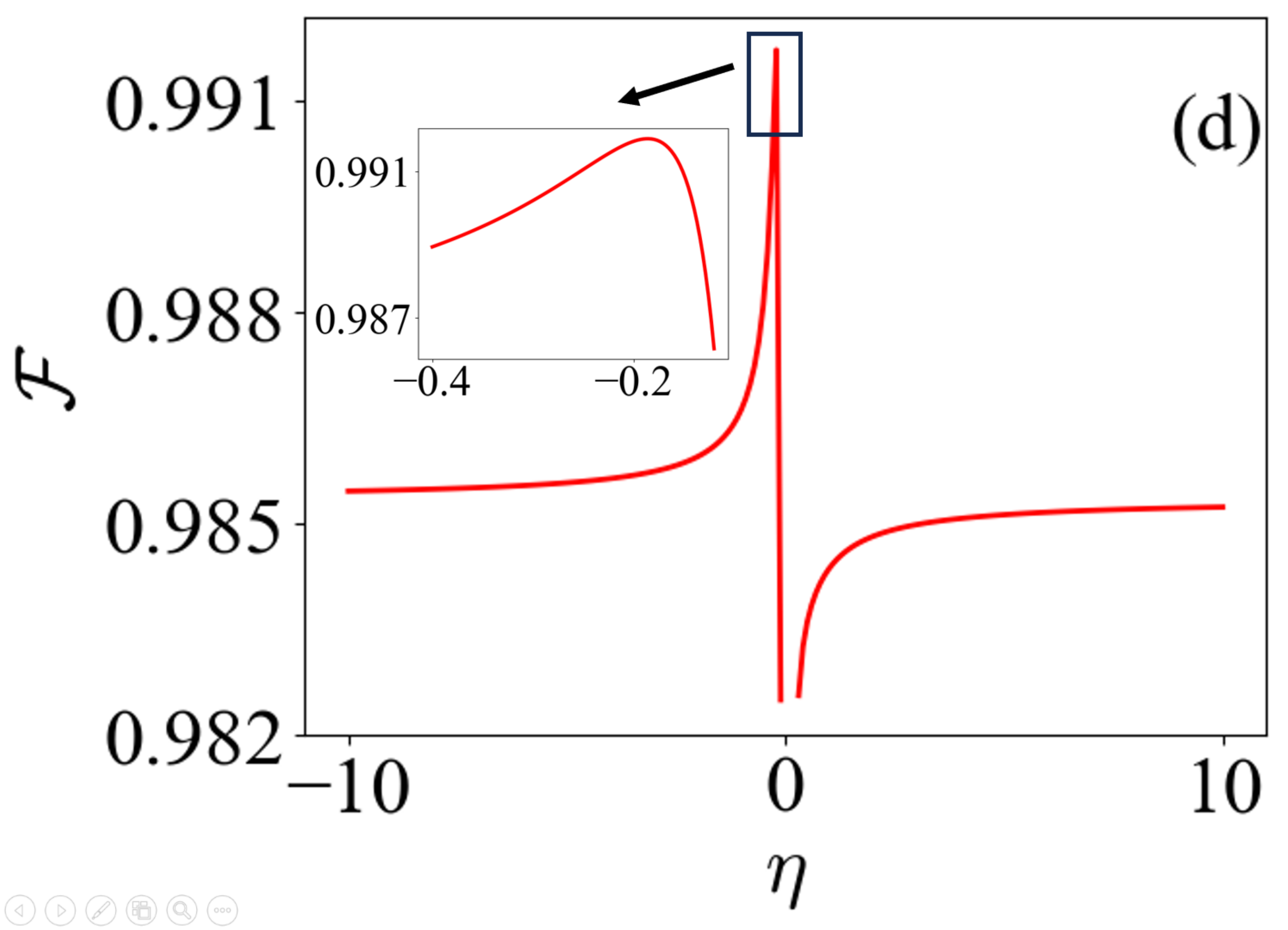}} 
   \end{minipage}
\begin{minipage}{0.23\textwidth}
	\centering
	\subfigure{\includegraphics[width=4.5cm]{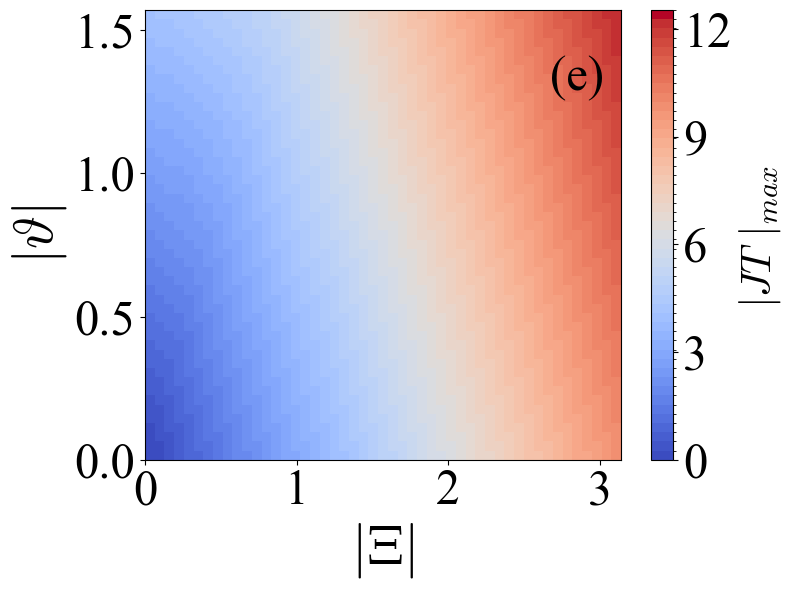}} 
\end{minipage}
\hfill 
\begin{minipage}{0.23\textwidth}
		\centering
		\subfigure{\includegraphics[width=4.0cm]{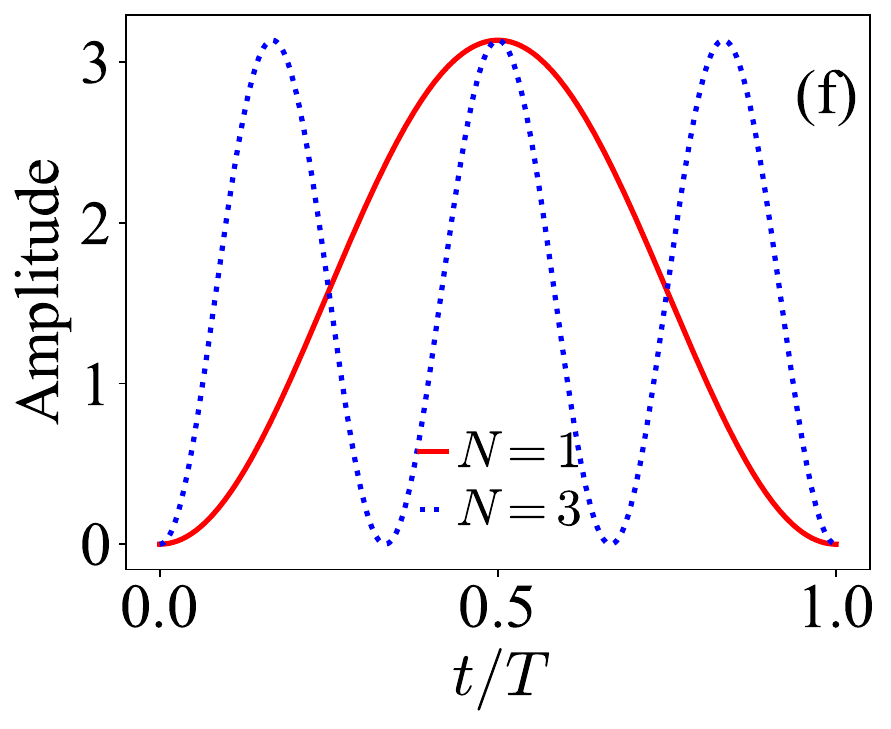}} 
	\end{minipage}
	   	\caption{(a) The fSim gate fidelity for 1600 different initial states under the rectangular pulses scheme, the red curve denotes the average fidelity, where gate time $T\approx 45$ ns and set $\delta E_z=2\pi/T\approx 2\pi\times 22$ MHz. (b) The $jT$ under the rectangular pulse scheme of fSim gate, where gate time $T\approx 45$ ns and set $\delta E_z=6\pi/T\approx 2\pi\times 66$ MHz. (c) The variation curve of error sensitivity $q_s$ with tunable parameter $\eta$, where the upper-left inset is a magnification of the boxed region. (d) The variation curve of fSim gate fidelity $\mathcal{F}$ with tunable parameter $\eta$,  where the upper-left inset showing a magnification of the boxed region. Here, each average fidelity value is obtained by averaging the results of numerical simulations performed over 100 distinct initial states. (e) The maximum value of pulse amplitude $|JT|_{max}$ with different fSim gate parameters $\{\vartheta, \Xi\}$ under the optimal parameter pulse scheme. Where the parameter ranges selected are $|\vartheta|\in[0,\pi/2]$ and $|\Xi|\in[0,\pi]$. (f) The $jT$ under the optimal parameter pulse scheme of fSim gate, where gate time $T\approx 50$ ns and set $\delta E_z=2N\pi/T\approx 2N\pi\times 20$ MHz. The red curve shows the result when $N=1$, and the blue curve shows the result when $N=3$. Where all the fSim gates parameters are chosen as $\vartheta=\pi/4,\Xi=\pi/2$.}
	   	\label{f4}
	   \end{figure}

	   In this section, we evaluate the performance of the quantum gates constructed using the method described above. Taking the fSim gate as a representative example, we combine numerical simulations with theoretical analyses to examine the theoretical fidelity under the RWA and decoherence, as well as the robustness of the gate against systematic errors—categorized as Rabi errors and detuning errors. Furthermore, we explore optimized gate construction by incorporating advanced quantum control techniques.To quantitatively assess gate performance, we adopt the average fidelity, defined as follows \cite{hx27}
	   \begin{align}\label{Eq.37}
	   	\mathcal{F}=\frac{1}{4\pi^2}\int_{0}^{2\pi}\int_{0}^{2\pi}|\langle\psi_f|\rho|\psi_f\rangle|^2d\varPhi_1d\varPhi_2,
	   \end{align}
	   where the initial state of the system is assumed to be pure state $|\psi_0\rangle=(\cos\varPhi_1|0\rangle_1+\sin\varPhi_1|1\rangle_1) (\cos\varPhi_2|0\rangle_2+\sin\varPhi_2|1\rangle_2) $,$|\psi_f\rangle$ is the ideal end state and $\rho$ is the density matrix of the actual end state of the system. Furthermore, to control the complexity of numerical simulations and reduce computational time, we fix the relative phase of the initial state to 0 and take 40 equally spaced values for each of $\varPhi_1$ and $\varPhi_2$  over the interval $[0,2\pi]$, replacing the aforementioned integral operation by averaging over a total of 1600 initial states.\\
	   \indent On the other hand, in the context of spin qubits in semiconductor QDs, charge noise stands out as the dominant noise source, which induces pure dephasing of the qubits. 
       To characterize the impact of qubit dephasing on gate fidelity, we can numerically solve the Lindblad master equation {\cite{hx64,hx65,hx66}}
	   \begin{align}\label{Eq.38}
	   	\dot{\rho}(t)=&i[\rho(t),H]-\sum_{j=1}^{2}\frac{\kappa_1}{2}(\sigma^\dagger_{j}\sigma_{j}\rho-2\sigma_{j}\rho\sigma^\dagger_{j}+\rho\sigma^\dagger_{j}\sigma_{j})\nonumber\\
	   	&-\sum_{j'=1}^{2}\frac{\kappa_2}{2}(\sigma^\dagger_{j'}\sigma_{j'}\rho-2\sigma_{j'}\rho\sigma^\dagger_{j'}+\rho\sigma^\dagger_{j'}\sigma_{j'}),
	   \end{align}
	   where $\sigma_{j}=|j\rangle\langle j|\otimes \widehat{I}, \sigma_{j'}= \widehat{I}\otimes|j'\rangle\langle j'|$, $\kappa_i=1/T_{2}^i$, $T_{2}^i$ represents the coherence time of the qubit $Q_i$. Here, we adopt the following experimental parameters \cite{hx67}: the coherence time of qubit $Q_1$ is  $T_2^1=120\ \mu s$, and the coherence time of qubit $Q_2$ is $T_2^2=61\ \mu$s.\\
	   \indent In the absence of experimental imperfections, the theoretical fidelity loss of the fSim gate originates solely from qubit decoherence and the neglection of high-frequency oscillatory terms ($j(t)e^{\pm i2\delta E_z t}$) by the rotating wave approximation. Decoherence-induced infidelity depends only on the gate evolution time, whereas the error introduced by the RWA is influenced by the frequency of these oscillatory terms. When we set 
	   $\delta E_z=2 \pi/T \approx 2\pi \times 22$ MHz, the numerical simulation results of the fSim gate fidelity for the rectangular pulse scheme are shown in Fig. \ref{f4}(a). We select 1600 different initial states for simulation, and the calculated average gate fidelity is $\mathcal{F}=98.56\%$, where the gate parameters are chosen as $(\vartheta=\pi/4,\Xi=\pi/2)$, and the gate time is $T\approx 45$ ns. \\
       \indent To reduce the RWA-induced error, the frequency of the oscillatory terms must be increased. We therefore consider $\delta E_z=2N\pi/T \approx 2N\pi \times 22$ MHz, $N=1,2,...,9$. Numerical simulation results show that when set $N=2$, the average fidelity rises to $\mathcal{F}=99.63\%$; when $N=3$, the average fidelity rises to $\mathcal{F}=99.82\%$. More results are presented in Table. \ref{tab1}. In addition, as $\delta E_z$ changes, the form of the rectangular pulse $j(t)$ has to change as well. According to the results of Eq. (\ref{Eq.23}), when $\delta E_z=2N\pi/T$, the change of $j(t)$ is equivalent to deflate the time axis to $1/N$ of the original one, and at the same time repeating it $N$ times. Taking $N=3$ as an example, the form of $j(t)$ is shown in Fig. \ref{f4}(b). It is worth noting that as $N$ increases, the portion of discontinuous leaps in the rectangular pulse scheme increases, which may introduce experimental challenges and additional errors.\\  
	   \indent To overcome the experimental difficulties of the rectangular pulse scheme, we consider a pulse of polynomial form $j(t)$ as follows
	
	   \begin{align}\label{Eq.39}
	   	j(t)=&\alpha[(1+2\beta+3\eta\beta)(\frac{t}{T})^2-(3\beta+4\eta\beta+2)(\frac{t}{T})^3\nonumber\\
	   	&+(\frac{t}{T})^4+\beta(\frac{t}{T})^5+\eta\beta(\frac{t}{T})^6].
	   \end{align}
	  For the sake of experimental implementation, we set the boundary conditions that $j(0)=j(T)=0$, $dj/dt|_{t=0}=dj/dt|_{t=T}$. Substituting these two boundary conditions into Eq. (\ref{Eq.23}), we get
	   \begin{align}\label{Eq.40}
	   	&\alpha=\frac{840\vartheta}{(14+35\beta+60\eta\beta)},\nonumber\\
	   	&\beta=\frac{2520(N\pi)^2-14(N\pi)^6(\frac{\Xi}{\vartheta})}{18900\eta+A(N\pi)^2+B(N\pi)^6(\frac{\Xi}{\vartheta})},\nonumber\\
	   	&A=-(6300+12600\eta),\quad B=60\eta+35.
	   \end{align}
   \renewcommand{\arraystretch}{1.3}
   \begin{table*}
   	\caption{The fSim gate fidelity and gate time for rectangular pulses and optimal parameter pulses are presented when $N$ takes different values. The gate fidelity is the average result of simulations for 1600 different initial states , and the gates parameters are chosen as $\vartheta=\pi/4,\Xi=\pi/2$. When $T=45$ ns, $\delta E_z=2N\pi/T\approx 2N\pi\times 22$ MHz; when $T=50$ ns, $\delta E_z=2N\pi/T = 2N\pi\times 20$ MHz. The experimentally achievable $\delta E_z=2\pi\times214$ MHz.}
   	\label{tab1} 
   	\begin{tabular}{cccccccccccc} 
   		\toprule 
   		\midrule
   		&$\delta E_z=2N\pi/T$& $N=1$ & $N=2$& $N=3$& $N=4$& $N=5$& $N=6$& $N=7$ & $N=8$& $N=9$ & $N=10$\\
   		\midrule 
   		\multirow{2}{*}{\parbox{2.7cm}{\centering Rectangular \\ Pulse Scheme}
   		}&$\mathcal{F}$ &$98.56\%$ &$99.63\%$ &  $99.82\%$ &$99.88\%$&$99.91\%$&$99.92\%$&$99.93\%$&$99.94\%$& $99.94\%$& $99.95\%$\\
   		&$T$&45 ns&45 ns&45 ns&45 ns&45 ns&45 ns&45 ns&45 ns &45 ns&$>$45 ns\\
   		\midrule
   		\multirow{2}{*}{\parbox{2.7cm}{\centering Optimal Parameter\\ Pulse Scheme }}&$\mathcal{F}$& $98.98\%$&$99.73\%$&$99.85\%$&$99.90\%$&$99.92\%$&$99.93\%$&$99.94\%$&$99.94\%$&$99.94\%$&$99.95\%$\\
   		&$T$&50 ns&50 ns&50 ns&50 ns&50 ns&50 ns&50 ns&50 ns&50 ns&50 ns\\
   		\midrule
   		\bottomrule 
   	\end{tabular}
   \end{table*}
	   In the following, we maintain $\delta E_z=2N\pi/T$. Note that once the fSim gate parameters $\{\vartheta,\Xi\}$ and the integer $N$ are determined, an adjustable parameter $\eta$ is still retained in $j(t)$. The purpose of retaining this parameter is to find the pulse scheme $j(\eta,t)$ with the best fidelity under approximation error through appropriate optimization methods, where $j(\eta,t)$ means that the form of $j(t)$ is determined by the tunable parameter $\eta$. \\
	   \indent To achieve this goal, we introduce the concept of error sensitivity \cite{hx68,hx69,hx70}, which is defined as follows: for a perturbative Hamiltonian quantity $H=H_0+\lambda\widetilde{H}$, where $\widetilde{H}$ is the perturbation term and $\lambda$ is a small quantity, the fidelity can be written up to the second order in the perturbation expansion as
	   \begin{align}\label{Eq.41}
	   	\mathcal{F}&=|\langle \psi(T)|\psi_0(T)\rangle|^2\nonumber\\
	   	&\approx 1-\lambda^2\left|\int_{0}^{T}dt\langle \psi_\perp(t)|\widetilde{H}|\psi_0(t)\rangle\right|^2,
	   \end{align}
	   $|\psi_0(t)\rangle$ is the ideal evolution state, and the ideal evolution matrix is defined as $U(s,t)=|\psi_0(s)\rangle\langle\psi_0(t)|+|\psi_\perp(s)\rangle\langle\psi_\perp(t)|$, for all the time $t$ satisfies $\langle\psi_\perp(t)|\psi_0(t)\rangle=0$, then the error sensitivity $q_s$ can be defined
	   \begin{align}\label{Eq.42}
	   	q_s&=-\frac{1}{2}\frac{\partial^2\mathcal{F}}{\partial \lambda^2}\vert_{\lambda=0}=-\frac{\partial\mathcal{F}}{\partial(\lambda^2)}\vert_{\lambda=0}\nonumber\\
	   	&=\left|\int_{0}^{T}dt\langle \psi_\perp(t)|\widetilde{H}|\psi_0(t)\rangle\right|^2.
	   \end{align}
       Now we calculate the error sensitivity $q_s$ under the approximation error caused by the RWA, since high-frequency oscillatory terms only exist in the $\{|01\rangle, |10\rangle\}$ subspace, the perturbation Hamiltonian within this subspace can be written as
	   \begin{align}\label{Eq.43}
	   	\widetilde{H}=\frac{1}{2}\left[\begin{array}{cc}
	   		0&j(t)e^{-2i\delta E_zt}\\
	   		j(t)e^{2i\delta E_zt}&0
	   	\end{array}\right],
	   \end{align}
	   and based on the evolution matrix in the $\{|01\rangle, |10\rangle\}$ subspace, $|\psi_0(t)\rangle$ and $|\psi_\perp(t)\rangle$ can be expressed as:
	   \begin{align}\label{Eq.44}
	   	&|\psi_0(t)\rangle=e^{-i\theta(t)}(\frac{\sqrt{2}}{2}|01\rangle+\frac{\sqrt{2}}{2}|10\rangle)\nonumber\\
	   	&|\psi_\perp(t)\rangle=e^{i\theta(t)}(\frac{\sqrt{2}}{2}|01\rangle-\frac{\sqrt{2}}{2}|10\rangle),
	   \end{align}
	   where $\theta=\int_{0}^{t}j(\tau)d\tau$. Substituting Eq. (\ref{Eq.43}) and Eq. (\ref{Eq.44})  into Eq. (\ref{Eq.42}) gives
	   \begin{align}\label{Eq.45}
	   	q_s=\left|\int_{0}^{T}dte^{-2i\theta(t)}j(t)\sin(2\delta E_z t)i/2\right|^2.
	   \end{align}
	   \indent In Fig. \ref{f4}(c), we numerically simulated the error sensitivity $q_s$ as a function of the adjustable parameter $\eta$. It should be noted that the error sensitivity $q_s$ is conventionally used for analyzing perturbative errors. Although the error term considered here is not a perturbation, $q_s$ is defined as an integral result, and the integral of the high-frequency oscillatory terms is a small quantity. Therefore, we argue that the error sensitivity can be used to analyze the approximation error induced by the RWA. To validate this approach, we also simulated the average fidelity of the fSim gate as a function of $\eta$, with results presented in Fig. \ref{f4}(d). A comparison between Fig. \ref{f4}(c) and Fig. \ref{f4}(d) reveals that the systematic error sensitivity and the average gate fidelity exhibit similar trends with respect to $\eta$, though the optimal values of $\eta$ differ slightly. Here we choose the parameter $\eta=-1/3$ corresponding to the minima of $q_s$ as the optimal parameter, and the polynomial-shaped pulse $j(t)|_{\eta=-\frac{1}{3} }$ is referred to as the optimal parameter pulse, because for different fSim gate parameters $\{\vartheta,\Xi\}$, the minima of $q_s$ are always at $\eta=-1/3$. \\
	   \indent The results for the maximum pulse amplitude $|JT|_{max}$ of the optimal parameter pulse scheme under different fSim gate parameters are shown in Fig. \ref{f4}(e). It can be observed that the overall pulse amplitude is larger than that of the rectangular pulse scheme, resulting in a longer gate time. According to the calculation results, the fSim gate time for any parameter under the optimal parameter pulse scheme will not exceed 100 ns, with an average gate time of 50 ns. Fig. \ref{f4}(f) shows a specific example of the pulse shape of $j(t)$, where the parameters of the fSim gate are $\vartheta=\pi/4, \Xi=\pi/2$, and $T\approx50$ ns. We also set $\delta E_z=2N\pi/T=2N\pi\times20$  MHz. When N=1, the shape of $j(t)$ is shown by the red curve in the figure. For $N=2,3,...,10$, it is only necessary to compress the horizontal axis to $1/N$ of its original length while repeating it $N$ times. The blue curve in the figure shows the result when $N=3$. As expected, the pulse variation under this scheme is smooth and continuous, thereby addressing the experimental challenges associated with abrupt transitions in the rectangular pulse approach.\\
	   \indent Finally, we give the average gate fidelity and gate evolution time of the rectangular pulse scheme and the optimal parametric pulse scheme for different $N$ values for $\delta E_z =2N\pi/T$ as shown in Table. \ref{tab1}. Here, the phases of the initial states used for calculating the average fidelity are all set to 0, and the results for different initial phases are presented in Appendix \ref{Appendix C}. It can be observed that for the same value of 
	   	$N$, the gate fidelity of the optimal parametric pulse scheme under the approximation error is slightly greater than that of the rectangular pulse scheme. In the presence of decoherence and approximation error, the theoretical fidelity of the fSim gate under the optimal parameter pulse scheme can reach up to $99.95\%$ with a gate time of 50 ns, where the fidelity loss mainly comes from the decoherence times of the qubits. With long coherence times of qubits (the coherence times of both qubit $Q_1$ and $Q_2$ are 100 ms), the theoretical fidelity of the fSim gate can exceed $99.99\%$. 
	   \subsection{Control errors}
	   In quantum state engineering, the main error sources are the random noise error due to the external environment and the control errors caused by the offset of the experimental control parameters. The systematic error can be divided into two parts, in which the offset of the pulse amplitude is called the Rabi error and the offset of the pulse frequency is called the detuning error. The total Hamiltonian with systematic error can be written in the form $H=H_0+\widetilde{H}$, where $H_0$ denotes the unperturbed original Hamiltonian and $\widetilde{H}$ denotes the perturbed  term.\\
	   \indent We first consider the robustness of the fSim gate under Rabi error. The original Hamiltonian quantity is given by Eq. (\ref{Eq.21}), where $B^R_y=B^L_y=0$ and the Rabi error is expressed as $J(t)\rightarrow (1+\Delta)J(t)$. Here, $\Delta$ denotes the magnitude of Rabi error, with its value range defining as $\Delta\in[-0.1,0.1]$. The perturbed term can be written as
	  \begin{align}\label{Eq.46}
	   \widetilde{H_\Delta}(t)=\Delta\left[\begin{array}{cccc}
	   		0&0&0&0\\
	   		0&-j\cos(\delta E_zt)&j(t)/2&0\\
	   		0&j(t)/2&-j\cos(\delta E_zt)&0\\
	   		0&0&0&0
	   	\end{array}\right].
	   \end{align}
	   \indent It can be observed that the perturbed term in Eq. (\ref{Eq.46}) commutes with the original Hamiltonian in Eq. (\ref{Eq.21}), and thus the total evolution matrix can be written as $U=U_0\widetilde{U}_\Delta$, where $U_0$ denotes the unperturbed total evolution matrix (i.e., the fSim gate). Thus, $\widetilde{U}_\Delta$ denotes the evolution matrix corresponding to the perturbed Hamiltonian, which is expressed as
	   
	  \begin{align}\label{Eq.47}
	   	\widetilde{U}_\Delta=\mathcal{T}e^{-i\int_{0}^{T}\widetilde{H_\Delta}(t)dt},
	   \end{align}
	   where $\mathcal{T}$ denotes the time-ordered product.For the perturbed term under the Rabi error, since it has nontrivial matrix elements only in the $\{|01\rangle, |10\rangle\}$ subspace, this perturbed term can be written within this subspace as
	   \begin{align}\label{Eq.48}
	   	\widetilde{H_\Delta}(t)=\Delta\frac{j(t)}{2}X-\Delta j(t)\cos(\delta E_z t)I,
	   \end{align}
	   in which $X,I$ are the Pauli matrix and the unit matrix in the $|01\rangle,|10\rangle$  basis. Since $\widetilde{H_\Delta}(t)$ at any two distinct times commutes, the time-ordered product in Eq. (\ref{Eq.47}) can be eliminated, and $\widetilde{U}_\delta$ results in
	   \begin{align}\label{Eq.49}
	   	\widetilde{U}_\Delta&=e^{-i\Delta\int_{0}^{T}j(t)\cos \delta E_ztdt}\nonumber\\
	   	&\times\left[\cos(\Delta \int_{0}^{T}\frac{j(t)}{2}dt)I-i\sin(\Delta\int_{0}^{T}\frac{j(t)}{2}dt)X\right]\nonumber\\
	   	&=e^{-i\delta\Xi}(\cos(\Delta\vartheta)I-i\sin(\Delta\vartheta)X).\nonumber\\
	   \end{align}
	   Substituting Eq. (\ref{Eq.49}) into Eq. (\ref{Eq.37}), the fidelity at this point can be solved and expressed as
	   \begin{align}\label{Eq.50}
	   	\mathcal{F}=\frac{1}{4\pi^2}\int_{0}^{2\pi}\int_{0}^{2\pi}|\langle\psi_0|\widetilde{U}_\Delta|\psi_0\rangle|^2d\varPhi_1d\varPhi_2.
	   \end{align}
	   \begin{figure}
	   \begin{minipage}{0.25\textwidth}
	   	\centering
	   	\subfigure{\includegraphics[width=4.5cm]{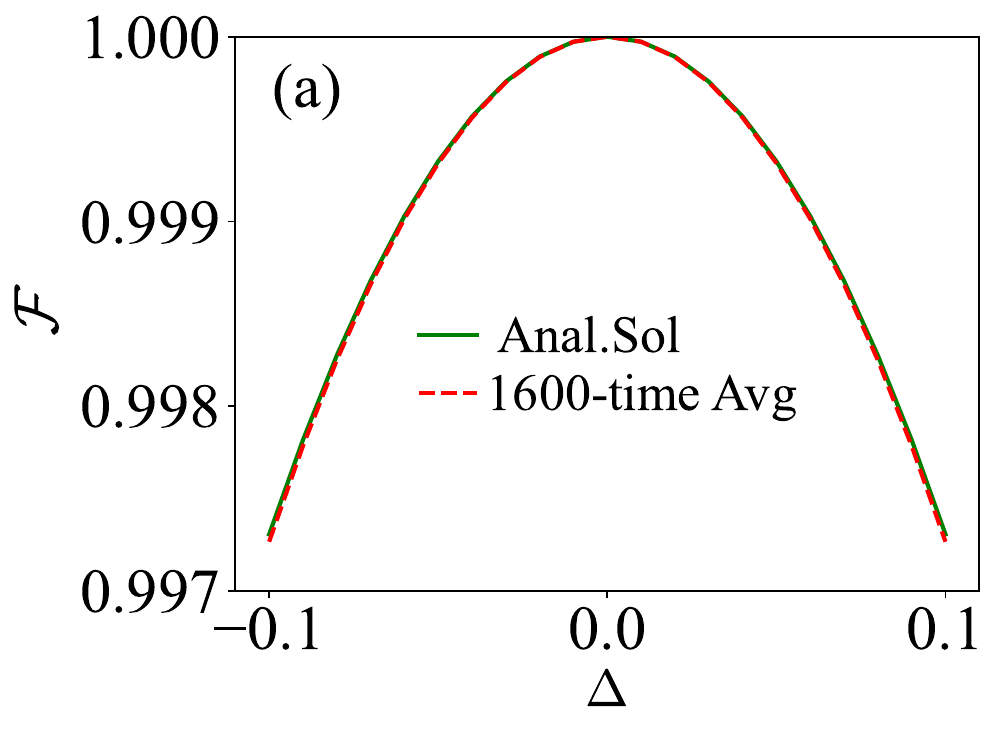}} 
	   \end{minipage}\begin{minipage}{0.24\textwidth}
	   	\centering
	   	\subfigure{\includegraphics[width=4.3cm]{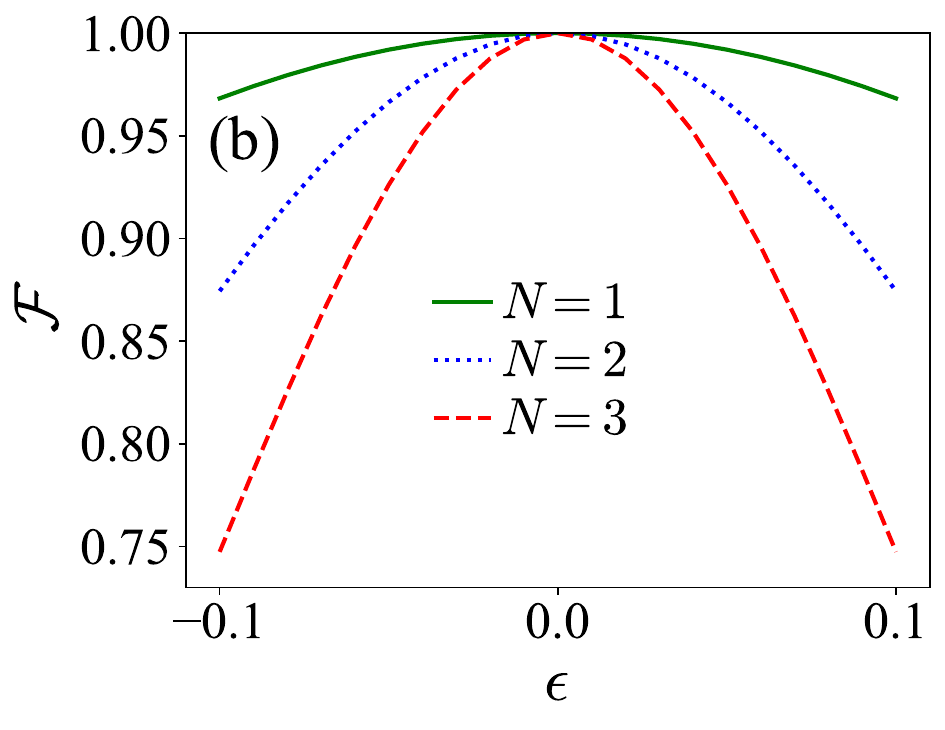}} 
	   \end{minipage}
	   \caption{(a) The variation curve of fSim gate fidelity with the magnitude of Rabi error $\Delta$, where the green curve is the result of analytical calculation, and the red curve is the average result of simulations with 1600 different initial states. (b) The variation curve of fSim gate fidelity with the magnitude of frequency detuning $\epsilon$ under the optimal parameter pulse. Where set $\delta E_z=2N\pi/T\approx 2N\pi\times 20$ MHz, the green curve corresponds to the result of $N=1$; the blue curve corresponds to $N=2$; the red curve corresponds to $N=3$. Where all the parameters of the fSim gates are chosen as $\vartheta=\pi/4,\Xi=\pi/2$.}
	   \label{f5}
	   \end{figure}
	   \indent Eqs. (\ref{Eq.49}) and (\ref{Eq.50}) show that the fidelity of the fSim gate under Rabi error depends only on the gate parameters  $\{\vartheta,\Xi\}$ and the magnitude of the Rabi error $\Delta$, while being independent of both the specific form of $j(t)$ and the magnitude of the $\delta E_z$. Here the gate parameters are still chosen as $\vartheta=\pi/4,\Xi=\pi/2$, and the analytic result for the fidelity of the fSim gate as a function of the Rabi error magnitude 
	   $\delta$ is obtained as
	   \begin{align}\label{Eq.51}
	   	\mathcal{F}=\frac{1}{32}[25+7\cos(\frac{\Delta\pi}{2})].
	   \end{align}
	   The variation curve of the fidelity $\mathcal{F}$ with the magnitude of the Rabi error $\Delta$ is shown in Fig. \ref{f5}(a), where we give both the analytical results of the theory and the numerical simulation results. The results show that the fSim gate has good stability under the Rabi error, and even for large amplitude fluctuations ($\Delta=\pm 0.1$), it still maintains a fidelity of more than $99.7\%$.\\
	   \indent We now turn to consider the robustness of the fSim gate under detuning error. In the silicon DQDs, frequency detuning is expressed as $\delta E_z\rightarrow(1+\epsilon_1)\delta E_z,E_z\rightarrow(1+\epsilon_2) E_z$, where $\epsilon$ denotes the magnitude of the detuning. For simplicity, here we assume that there is a deviation of the same magnitude $(\epsilon_1=\epsilon_2=\epsilon)$ at both frequencies, and similarly specify that the range of its values is $\epsilon\in[-0.1,0.1]$.The perturbed  term can be written as
	   \begin{align}\label{Eq.52}
	\widetilde{H_\epsilon}(t)=\epsilon\left[\begin{array}{cccc}
	   		E_z&0&0&0\\
	   		0&-\delta E_z/2&0&0\\
	   		0&0&\delta E_z/2&0\\
	   		0&0&0&-E_z
	   	\end{array}\right].
	   \end{align}
	   Since the perturbed  term does not commute with the original Hamiltonian in Eq. (\ref{Eq.21}), here we only give numerical simulation results for the fidelity of the fSim gate under detuning error, and it can be observed that the magnitude of the detuning error is related to the value of 
	   $\delta E_z$. We set $\delta E_z=2N\pi/T\approx 2N\pi\times 20$ MHz and present the numerical simulation results for the fidelity of the fSim gate constructed by the optimal parametric pulse scheme when $N=1,2,3$, as shown in Fig. \ref{f5}(b). The results show that when $\delta E_z$ is small ($N=1$), the fSim gate exhibits good stability under detuning error. As $\delta E_z$ increases, the infidelity caused by frequency detuning grows rapidly and soon dominates over the approximation error introduced by the RWA. This suggests a trade-off in selecting the parameter $N$, for small detuning, increasing $N$ helps suppress RWA-related errors and improves fidelity; for large detuning, a smaller value of $N$ is preferable to mitigate the more significant detuning-induced infidelity.
	   \section{The geometric+dynamical fSim gate\label{sec5}}
	   	\indent In recent years, various strategies have been proposed to mitigate quantum gate errors, among which geometric quantum gates have attracted notable attention due to their potential for high-fidelity operation. Unlike conventional dynamic methods that rely on dynamic phases, geometric quantum gates are implemented using purely geometric phases, which depend only on the global evolution path of the system. Owing to their global nature, these gates can inherently suppress errors induced by fluctuations in control Hamiltonian parameters. As a result, geometric-phase-based gates are considered a promising approach for achieving robustness against control inaccuracies \cite{hx71,hx72}. In this section, we introduce a hybrid geometric–dynamical scheme for constructing the fSim gate by integrating geometric quantum gate principles. Through numerical simulations, we demonstrate that this scheme exhibits significantly improved robustness to systematic errors compared to purely dynamical implementations. \\
	   	\indent For constructing fSim gates that utilize geometric phases, some reconstructions of the original Hamiltonian quantity are required.  We start from Eq. (\ref{Eq.19}) and set $B_y^R(t)=B_y^L(t)=0,J=2j\cos(\omega t+\psi)$ and $\delta E_z=\omega$. Performing a rotating frame transformation of Eq. (\ref{Eq.19}) with a rotation matrix of $U''_t=\mathrm{exp}(-i(\omega t/4)\sigma^1_z\otimes I+i(\omega t/4)I\otimes\sigma^2_z)$, the Hamiltonian under the rotating wave approximation, in the new basis of $\{|\widetilde{00}\rangle, |\widetilde{01}\rangle, |\widetilde{10}\rangle, |\widetilde{11}\rangle\}$, is written  as 
	   	\begin{align}\label{Eq.53}
	   		H_g=\left[\begin{array}{cccc}
	   			E_z+j\cos(\omega t)&0&0&0\\
	   			0&0&je^{i\psi}/2&0\\
	   			0&je^{-i\psi}/2&0&0\\
	   			0&0&0&j\cos(\omega t)-E_z
	   		\end{array}\right].
	   	\end{align}
	   	 It can be readily observed that Hamiltonian exhibits no coupling between the $|\widetilde{01}\rangle, |\widetilde{10}\rangle$ energy levels and the $|\widetilde{00}\rangle, |\widetilde{11}\rangle$ energy levels, and thus these two sets of energy levels can be analyzed separately. Our goal is to construct a pure geometric phase iSWAP-like gate in the in the $\{|\widetilde{01}\rangle, |\widetilde{10}\rangle\}$ subspace, while constructing a dynamical CPHASE gate in the $\{|\widetilde{00}\rangle, |\widetilde{11}\rangle\}$ subspace, thereby obtaining a geometric+dynamical fSim gate.\\
	   	\indent In $\{|\widetilde{01}\rangle, |\widetilde{10}\rangle\}$ subspace, the Hamiltonian is $H_{c}=\Omega|\widetilde{01}\rangle\langle \widetilde{10}|+H.c.$, where $\Omega=je^{i\psi}/2$. To construct a pure geometric iSWAP-like gate \cite{hx30,hx73} , we set 
	   	\begin{align}\label{Eq.54}
	   		&\Omega=(j(t)/2)e^{-\frac{\pi}{2}i},\quad\ \ \ \ 0<t\le T_1,\nonumber\\
	   		&\Omega=(j(t)/2)e^{(\frac{\pi}{2}-\vartheta)i},\quad T_1<t\le T_2,\nonumber\\
	   		&\Omega=(j(t)/2)e^{-\frac{\pi}{2}i},\quad\ \ \ \ T_2<t\le T,
	   	\end{align}
	   	and 
	   	\begin{align}\label{Eq.55}
	   		\int_{0}^{T_1}j(t)dt=\frac{\pi}{4},\quad \int_{T_1}^{T_2}j(t)dt=\frac{\pi}{2},\quad \int_{T_2}^{T}j(t)dt=\frac{\pi}{4}.
	   	\end{align}
	   	Meanwhile, the evolution matrix of the system can be written as
	   	\begin{align}\label{Eq.56}
	   		U(T)&=U_3(T,T_2)U_2(T_2,T_1)U_1(T_1,0)\nonumber\\
	   		&=e^{\int_{T_2}^{T}H_{c}^1(t)dt}e^{\int_{T_1}^{T_2}H_c^{2}(t)}dte^{\int_{0}^{T_1}H_{c}^1(t)dt},
	   	\end{align}
	   	where
	   	\begin{align}\label{Eq.57}
	   		&H_{c}^1(t)=(j(t)/2)e^{-\frac{\pi}{2}i}|\widetilde{01}\rangle\langle \widetilde{10}|+H.c,\nonumber\\
	   		&H_{c}^2(t)=(j(t)/2)e^{(\frac{\pi}{2}-\vartheta)i}|\widetilde{01}\rangle\langle \widetilde{10}|+H.c..
	   	\end{align}
	   	As a result, we can obtain
	   	\begin{align}\label{Eq.58}
	   		U(T)=\left[\begin{array}{cc}
	   			\cos \vartheta&-i\sin\vartheta\\
	   			-i\sin\vartheta&\cos\vartheta
	   		\end{array}\right],
	   	\end{align}
	   	this constitutes an iSWAP-like gate. To prove that this is a geometric quantum gate, we define a set of orthogonal states
	   	\begin{align}\label{Eq.59}
	   		|b\rangle=\frac{\sqrt{2}}{2}|\widetilde{01}\rangle-\frac{\sqrt{2}}{2}|\widetilde{10}\rangle,\nonumber\\
	   		|d\rangle=\frac{\sqrt{2}}{2}|\widetilde{01}\rangle+\frac{\sqrt{2}}{2}|\widetilde{10}\rangle,
	   	\end{align}
	   	Eq. (\ref{Eq.58}) can be written as
	   	\begin{align}\label{Eq.60}
	   		U(T)=e^{i\vartheta}|b\rangle\langle b|+e^{-i\vartheta}|d\rangle\langle d|.
	   	\end{align}
   	  \begin{figure}
   	  	\begin{minipage}{0.24\textwidth}
   	  		\centering
   	  		\subfigure{\includegraphics[width=3.5cm]{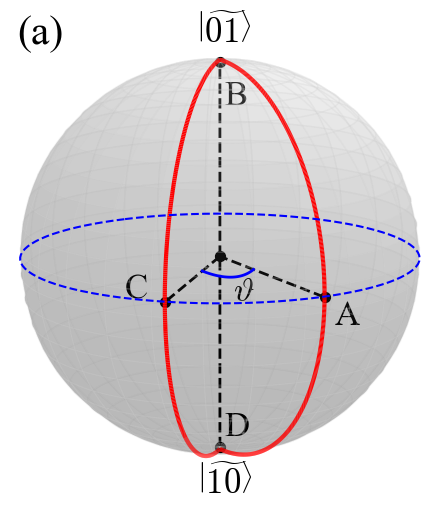}} 
   	  	\end{minipage}\begin{minipage}{0.24\textwidth}
   	  		\centering
   	  		\subfigure{\includegraphics[width=4.3cm]{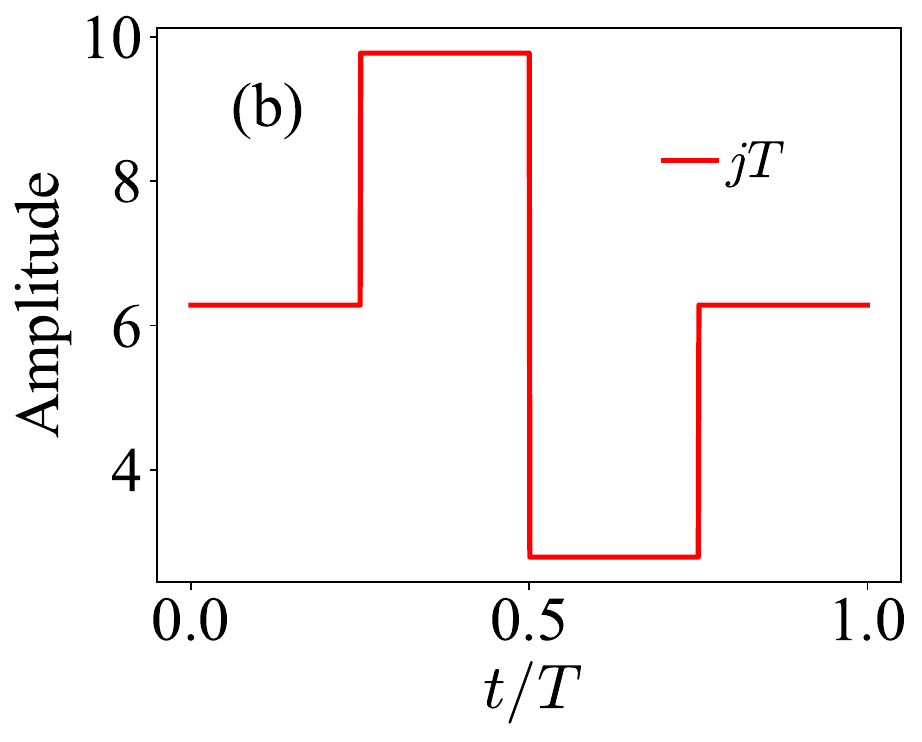}} 
   	  	\end{minipage}
     	\begin{minipage}{0.24\textwidth}
     		\centering
     		\subfigure{\includegraphics[width=4.3cm]{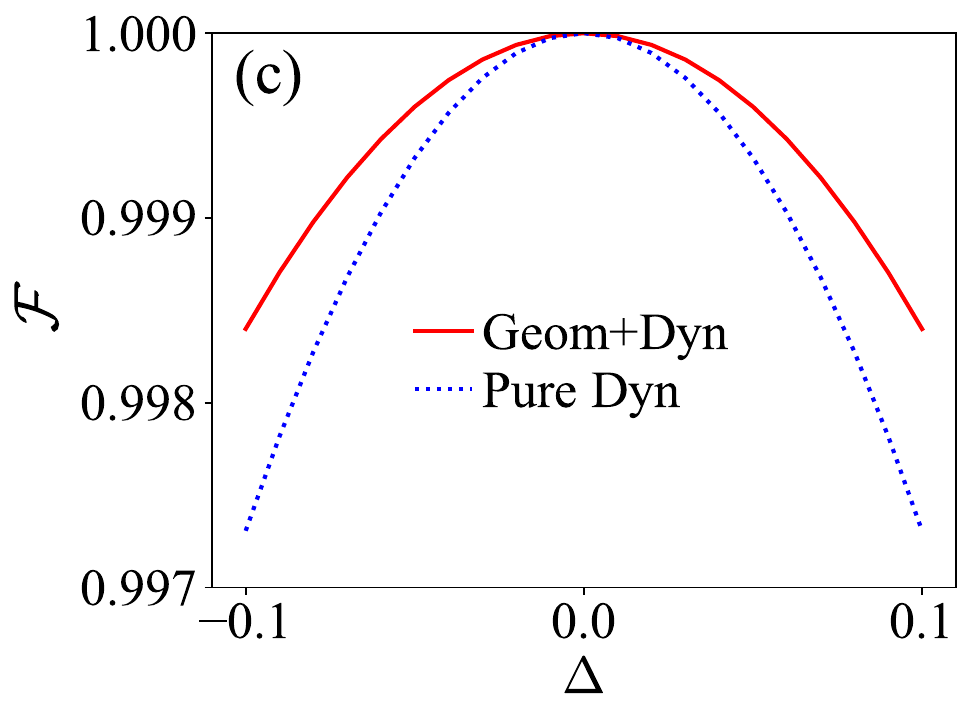}} 
     	\end{minipage}\begin{minipage}{0.24\textwidth}
     		\centering
     		\subfigure{\includegraphics[width=4.3cm]{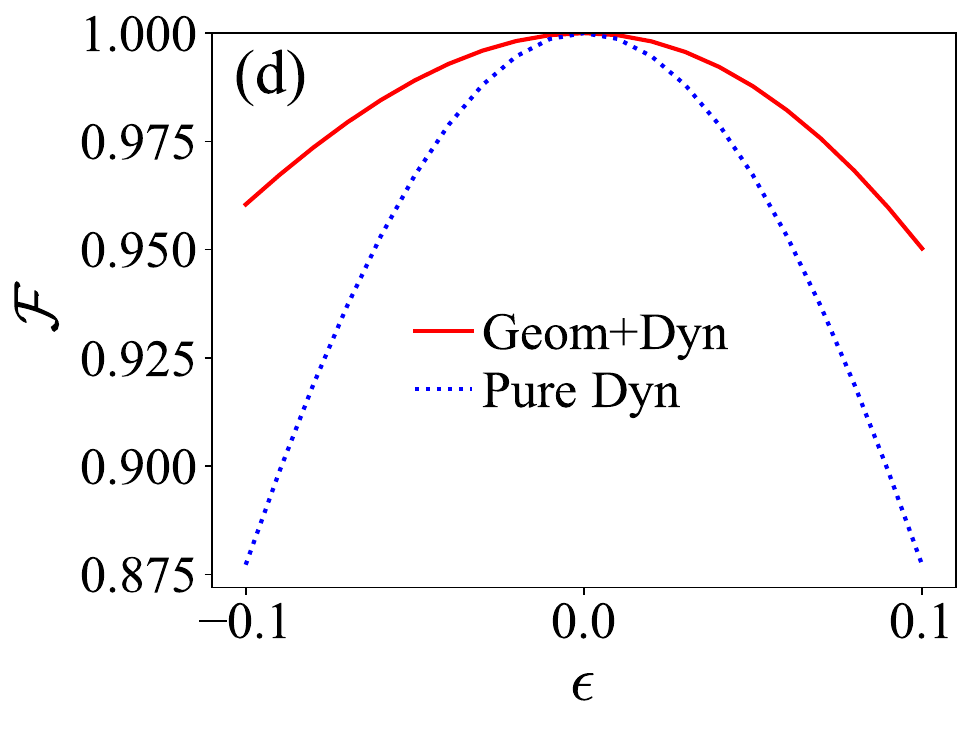}} 
     	\end{minipage}
   	  	\caption{(a) The evolution trajectory of $|b(t)\rangle$ on the Bloch sphere, with the path A-B-C-D-A. (b) The $jT$ under the geometric-dynamic fSim gate scheme, where $T\approx 158$ ns and set $\delta E_z=4\pi/T\approx 2\pi \times 6.3$ MHz. (c) The variation curves of fSim gate fidelity for the geometric+dynamic scheme and pure dynamic scheme with the magnitude of Rabi error, where the red curve represents the geometric-dynamic scheme and the blue curve represents the pure dynamic scheme. (d) The variation curves of fSim gate fidelity for the geometric+dynamic scheme and pure dynamic scheme with the magnitude of frequency detuning error, where the red curve represents the geometric-dynamic scheme and the blue curve represents the pure dynamic scheme, where both scheme set $\delta E_z=4\pi/T$. Where all the parameters of the fSim gate are chosen as $\vartheta=\pi/4,\Xi=\pi/2$.}
   	  	\label{f6}
   	  \end{figure}
	   	The evolution trajectory of $|b(t)\rangle=U(t,0)|b\rangle$ on the Bloch sphere constructed by the $\{|\widetilde{01}\rangle,|\widetilde{10}\rangle\}$ subspace is shown in Fig. \ref{f6}(a). From initial time 0 to the time $T_1$, it starts from point A and evolves along a meridian to the north pole B; from $T_1$ to $T_2$, it evolves along another meridian through B-C-D to the south pole, where the angle between the two meridians is $\vartheta$. Finally, from $T_2$ to $T$, it returns from D to A along the first meridian.  Such that $|b(t)\rangle$ undergoes a cyclic evolution from 0 to $T$, differing only by a global phase $e^{i\vartheta}$ before and after the evolution. Additionally, the parallel transport condition is always satisfied throughout the whole evolution \cite{hx74,hx75}:
	   	\begin{align}\label{Eq.61}
	   		&\langle b(t)|H_{c}^1|b(t)\rangle=0,\quad\ \ \ \ 0<t\le T_1,\nonumber\\
	   		&\langle b(t)|H_{c}^2|b(t)\rangle=0,\quad\ \ \ \ T_1<t\le T_2,\nonumber\\
	   		&\langle b(t)|H_{c}^1|b(t)\rangle=0,\quad\ \ \ \ T_2<t\le T.
	   	\end{align}
	   	Subsequently, the dynamic phase during the evolution is always zero, and the total phase difference before and after the evolution originates only from the geometric phase. It can be shown that the above conclusion still holds for $|d(t)\rangle$. Therefore, this constitutes a geometric iSWAP-like gate. \\
	   	\indent Then we consider the time evolution of system in the  $\{|\widetilde{00}\rangle,|\widetilde{11}\rangle\}$ subspace. Since the sub-Hamiltonian has nontrivial components only on the diagonal elements, a geometric construction scheme is not found, where the CPHASE gate part of the fSim gate is still constructed using the same dynamics scheme as in the Eq. (\ref{Eq.23}). In summary, the conditions that all physical quantities required to construct a geometric+dynamical fSim gate must satisfy are
	   	\begin{align}\label{Eq.62}
	   		&J=2j\cos(\omega t+\pi/2),\quad\ \ \ \ \ \ \   t\in [0,T_1)\cup[T_2,T],\nonumber \\
	   		&J=2j\cos(\omega t+\vartheta-\pi/2), \quad t\in [T_1,T_2),\nonumber\\
	   		&\int_{0}^{T_1}j(t)dt=\frac{\pi}{2},\quad \int_{T_1}^{T_2}j(t)dt=\frac{\pi}{4},\quad \int_{T_2}^{T}j(t)dt=\frac{\pi}{2},\nonumber\\
	   		&\int_{0}^{T}J(t)=-\Xi, \quad E_z=\frac{\Xi}{2T},\quad \delta E_z=\omega.
	   	\end{align}
	   	Since the geometric gate scheme requires discontinuous phase switching of the pulses, we present here a specific construction scheme for the fSim gate where that $j(t)$ takes a simple rectangular form
	   	\begin{align}\label{Eq.63}
	   		&j=\frac{2\pi}{T},\quad\quad\quad \quad \quad \ \ \ \psi=\frac{\pi}{2}, \quad \quad\quad 0<t\le \frac{T}{4},\nonumber\\
	   		&j=\frac{4\pi\cos\vartheta+\pi\Xi}{2T\cos\vartheta},\quad \psi=\vartheta-\frac{\pi}{2}, \quad \frac{T}{4}<t\le \frac{T}{2},\nonumber\\
	   		&j=\frac{4\pi\cos\vartheta-\pi\Xi}{2T\cos\vartheta},\quad \psi=\vartheta-\frac{\pi}{2}, \quad \frac{T}{2}<t\le \frac{3T}{4},\nonumber\\
	   		&j=\frac{2\pi}{T},\quad \quad\quad \quad \quad\ \ \ \psi=\frac{\pi}{2},\quad \quad \ \ \ \frac{3T}{4}<t\le T,\nonumber\\
	   		&E_z=\frac{\Xi}{2T},\quad \ \ \ \ \delta E_z=4\pi/T.
	   	\end{align}
	    When fSim gate parameters are still chosen as $\vartheta=\pi/4,\Xi=\pi/2$, the impulse $jT$ is of the form shown in Fig. \ref{f6}(b). After that, we compare the stability of the fSim gate of the geometric+dynamics scheme with that of the pure dynamics scheme under the control errors, and give out the curves of the gate fidelity of the two schemes with the size of the control error by means of numerical simulation are shown in Fig. \ref{f6}(c) and (d). For comparison, the systematic error here is defined in the same way as in Sec. \ref{sec4}, and $\delta E_z=4\pi/T$ is uniformly selected for both. It can be observed that under both Rabi error and detuning error, the fSim gate based on the geometric+dynamical scheme exhibits better robustness than that based on the pure dynamical scheme. In addition, it should be noted that although the geometric scheme shows good results in theory, it requires the state vector to undergo cyclic evolution on the Bloch sphere, which leads to a longer gate evolution time and increases the loss of fidelity caused by decoherence. 
	   	\section{Summary\label{sec6}}
	   	\indent In summary, we propose a novel method for constructing robust two-qubit fSim and B gates, whose core innovation lies in the simultaneous handling of state transfer processes across multiple energy levels, demonstrating significant advantages in the construction of composite two-qubit gates. Theoretically, this approach enables the derivation of a parameterized gate set that can be directly implemented in any two-qubit physical system, with the types of gates determined by the symmetry of the Hamiltonian. Applied to silicon DQDs, we demonstrate a one-step fSim gate scheme and a B gate scheme requiring only a single pulse switch. Furthermore, the method can be naturally integrated with various optimization strategies to enhance gate performance: combining it with quantum optimal control, we develop an optimized parametric pulse scheme that improves gate fidelity under approximation errors; incorporating geometric quantum gate theory, we construct a geometric–dynamical scheme that significantly enhances robustness against systematic errors.\\
	   	\indent Our method is highly versatile and can be readily adapted to various physical platforms for two-qubit systems. Depending on the symmetries inherent to a given physical implementation system, distinct parameterized families of two-qubit gates can be constructed. By appropriately choosing the parameters, complex composite two-qubit gates can be implemented using a minimal number of control operations. Moreover, the method imposes constraints only on the time-integrated effect of the control pulses, not on their specific waveforms, thereby exhibiting excellent compatibility with a wide range of pulse-shaping and optimal control frameworks. Future work can leverage various optimization theories to develop fast, robust implementations of composite gates under different experimental error models, further enhancing the practicality and scalability of this approach.
	   	\section*{ACKNOWLEDGMENTS}
	   	The authors acknowledge the financial support by National Natural Science Foundation of China~(Grants  No. 62471001, No. 12475009, and No. 12075001), Natural Science Research Project in Universities of Anhui Province (No. 2024AH050068), Anhui Provincial Key Research and Development Plan (Grant No. 2022b13020004), Anhui Province Science and Technology Innovation Project (Grant No. 202423r06050004).
	   	\section*{DATA AVAILABILITY}
	   The data that support the findings of this paper are openly available \cite{hx76}.
	   	\onecolumngrid 
		\appendix
		\renewcommand{\appendixname}{APPENDIX}  
		\section{CONSTRUCTING UNIVERSAL TWO-QUBIT GATES USING B GATE\label{Appendix A}}
\begin{figure}
		 	\begin{minipage}{0.6\textwidth}
		 		\centering
		 		\subfigure{\includegraphics[width=10cm]{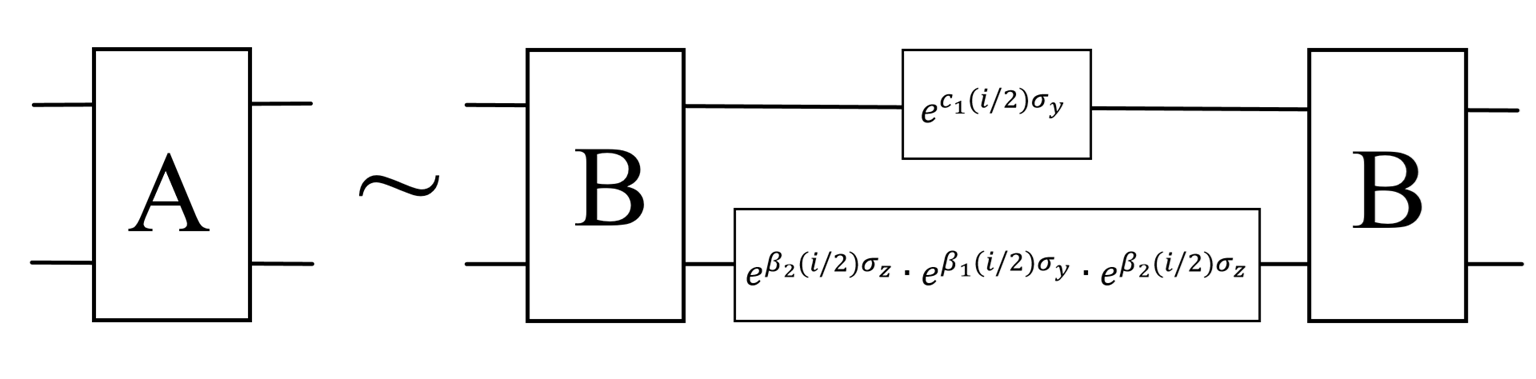}} 
		 	\end{minipage}
		 	\caption{By using B gates and single-qubit gates to construct arbitrary nonlocal two-qubit gate A.}
		 	\label{FA1}
		 \end{figure}

		According to Cartan's KAK Decomposition, an arbitrary two-qubit unitary gate $U\in SU(4)$ can be written as:
		\begin{align}\label{a1}
		U=(k_1\otimes k_2)A(c_1,c_2,c_3)(k_3\otimes k_4)
		\end{align}
		where $k_i\in SU(2)$ is single-qubite gate, while
		\begin{align}\label{a2}
			A=e^{(i/2)c_1\sigma_x^1\sigma_x^2}e^{(i/2)c_2\sigma_y^1\sigma_y^2}e^{(i/2)c_3\sigma_z^1\sigma_z^2}
		\end{align} 
		 is the nonlocal part, determined by three real parameters $c_1,c_2$ and $c_3$. \\
		 \indent The B gate can implement an arbitrary two-qubit unitary operation with the minimum number of two-qubit gates and single-qubit gates. A nonlocal A gate can be constructed merely by applying two B gates and two single-qubit gates, with its specific combination form as shown in Fig. \ref{FA1}, where the parameters $\beta_1$, $\beta_2$ are satisfied
		\begin{align}\label{a3}
			&\cos\beta_1=1-4\sin^2\frac{c_2}{2}\cos^2\frac{c_3}{2},\nonumber\\
			&\sin\beta_2=\sqrt{\frac{\cos c_2\cos c_3}{1-2\sin^2\frac{c_2}{2}\cos^2\frac{c_3}{2}}} .
		\end{align}
		Therefore, implementing an arbitrary two-qubit unitary gate only applying two B gates and six single-qubit gates.
		\section{PARAMETERIZED TOTAL HAMILTONIAN\label{Appendix B}}
		The parameterized total Hamiltonian can be written as
		\begin{align}\label{Eq.B.1}
			H(t)=\hbar\left[\begin{array}{cccc}
				0& i\Omega_{12}(t)e^{-i\varphi_2}&i\Omega_{13}(t)e^{-i\varphi_3}&i\Omega_{14}(t)e^{-i\varphi_4}\\
				-i\Omega_{12}(t)e^{i\varphi_2}&-\dot{\varphi}_2&i \Omega_{23}(t)e^{i(\varphi_2-\varphi_3)}&i\Omega_{24}(t)e^{i(\varphi_2-\varphi_4)}\\
				-i\Omega_{13}(t)e^{i\varphi_3}&-i\Omega_{23}(t)e^{-i(\varphi_2-\varphi_3)} & -\dot{\varphi}_3&i\Omega_{34}(t)e^{i(\varphi_3-\varphi_4)}\\
				-i\Omega_{14}(t)e^{i\varphi_4}&-i\Omega_{24}(t)e^{-i(\varphi_2-\varphi_4)}&-i\Omega_{34}(t)e^{-i(\varphi_3-\varphi_4)}&-\dot{\varphi}_4
			\end{array}\right],
		\end{align}
		where
		\begin{align}\label{Eq.B.2}
			\Omega_{12}&=\sin \gamma_1 \sin \theta_1 (\dot{\theta}_1 \cos \gamma_1 - \dot{\phi}_1 \sin \gamma_1 \sin \theta_1)+ \sin \gamma_2 \sin \theta_2 (\dot{\theta}_2 \cos \gamma_2 + \dot{\phi}_2 \sin \gamma_2 \sin \theta_2) \nonumber\\&- \dot{\gamma}_1 \cos \theta_1 - \dot{\gamma}_2 \cos \theta_2,\nonumber\\
			\Omega_{13}&=\dot{\theta}_1 \sin \gamma_1 (\sin \gamma_1 \sin \phi_1 - \cos \gamma_1 \cos \theta_1 \cos \phi_1) - \dot{\theta}_2 \sin \gamma_2 (\sin \gamma_2 \sin \phi_2 + \cos \gamma_2 \cos \theta_2 \cos \phi_2)\nonumber\\
			&- \dot{\gamma}_1 \sin \theta_1 \cos \phi_1 - \dot{\gamma}_2 \sin \theta_2 \cos \phi_2 + \dot{\phi}_1 \sin \gamma_1 \sin \theta_1 (\cos \gamma_1 \sin \phi_1 + \sin \gamma_1 \cos \theta_1 \cos \phi_1)\nonumber\\
			&+ \dot{\phi}_2 \sin \gamma_2 \sin \theta_2 (\cos \gamma_2 \sin \phi_2 - \sin \gamma_2 \cos \theta_2 \cos \phi_2),\nonumber\\
			\Omega_{14}&=-\dot{\theta}_1 \sin \gamma_1 (\sin \gamma_1 \cos \phi_1 + \cos \gamma_1 \cos \theta_1 \sin \phi_1) + \dot{\theta}_2 \sin \gamma_2 (\sin \gamma_2 \cos \phi_2 - \cos \gamma_2 \cos \theta_2 \sin \phi_2)  \nonumber\\&- \dot{\gamma}_1 \sin \theta_1 \sin \phi_1- \dot{\gamma}_2 \sin \theta_2 \sin \phi_2 - \dot{\phi}_1 \sin \gamma_1 \sin \theta_1 (\cos \gamma_1 \cos \phi_1 - \sin \gamma_1 \cos \theta_1 \sin \phi_1) \nonumber\\
			&- \dot{\phi}_2 \sin \gamma_2 \sin \theta_2 (\cos \gamma_2 \cos \phi_2 + \sin \gamma_2 \cos \theta_2 \sin \phi_2),\nonumber\\
			\Omega_{23}&= -\dot{\theta}_1 \sin \gamma_1 (\sin \gamma_1 \cos \phi_1 + \cos \gamma_1 \cos \theta_1 \sin \phi_1) - \dot{\theta}_2 \sin \gamma_2 (\sin \gamma_2 \cos \phi_2 - \cos \gamma_2 \cos \theta_2 \sin \phi_2),\nonumber\\
			&- \dot{\gamma}_1 \sin \theta_1 \sin \phi_1 + \dot{\gamma}_2 \sin \theta_2 \sin \phi_2 - \dot{\phi}_1 \sin \gamma_1 \sin \theta_1 (\cos \gamma_1 \cos \phi_1 - \sin \gamma_1 \cos \theta_1 \sin \phi_1) \nonumber\\
			&+ \dot{\phi}_2 \sin \gamma_2 \sin \theta_2 (\cos \gamma_2 \cos \phi_2 + \sin \gamma_2 \cos \theta_2 \sin \phi_2),\nonumber\\
			\Omega_{24}&=-\dot{\theta}_1 \sin \gamma_1 (\sin \gamma_1 \sin \phi_1 - \cos \gamma_1 \cos \theta_1 \cos \phi_1) - \dot{\theta}_2 \sin \gamma_2 (\sin \gamma_2 \sin \phi_2 + \cos \gamma_2 \cos \theta_2 \cos \phi_2)\nonumber\\
			&+ \dot{\gamma}_1 \sin \theta_1 \cos \phi_1 - \dot{\gamma}_2 \sin \theta_2 \cos \phi_2 - \dot{\phi}_1 \sin \gamma_1 \sin \theta_1 (\cos \gamma_1 \sin \phi_1 + \sin \gamma_1 \cos \theta_1 \cos \phi_1)\nonumber\\
			&+ \dot{\phi}_2 \sin \gamma_2 \sin \theta_2 (\cos \gamma_2 \sin \phi_2 - \sin \gamma_2 \cos \theta_2 \cos \phi_2),\nonumber\\
			\Omega_{23}&=\sin \gamma_1 \sin \theta_1 (\dot{\theta}_1 \cos \gamma_1 - \dot{\phi}_1 \sin \gamma_1 \sin \theta_1) - \sin \gamma_2 \sin \theta_2 (\dot{\theta}_2 \cos \gamma_2 + \dot{\phi}_2 \sin \gamma_2 \sin \theta_2) \nonumber\\
			&- \dot{\gamma}_1 \cos \theta_1 + \dot{\gamma}_2 \cos \theta_2 .
		\end{align}
\begin{figure}
	\begin{minipage}{0.49\textwidth}
		\centering
		\subfigure{\includegraphics[width=8cm]{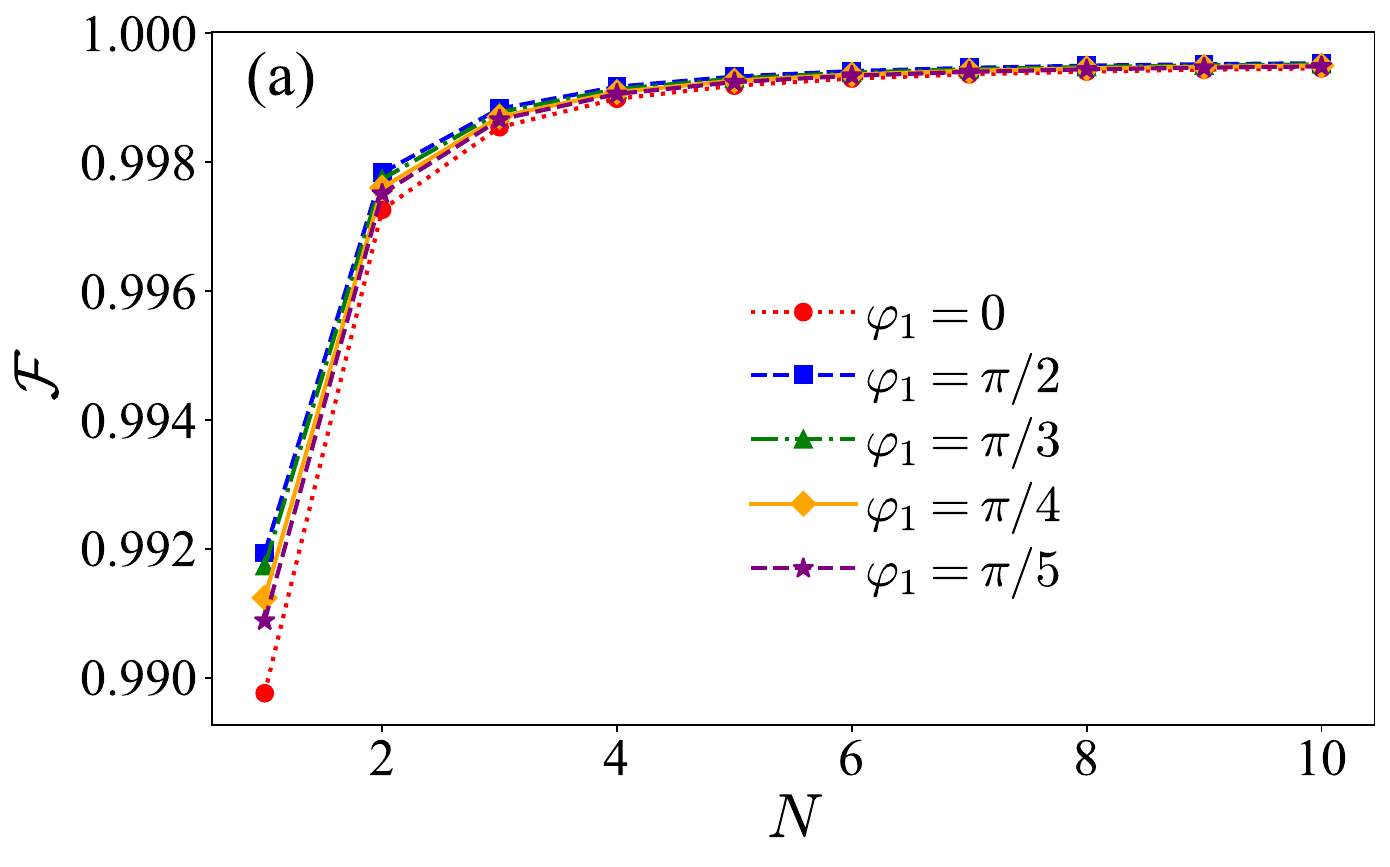}} 
	\end{minipage}
	\begin{minipage}{0.49\textwidth}
		\centering
		\subfigure{\includegraphics[width=8cm]{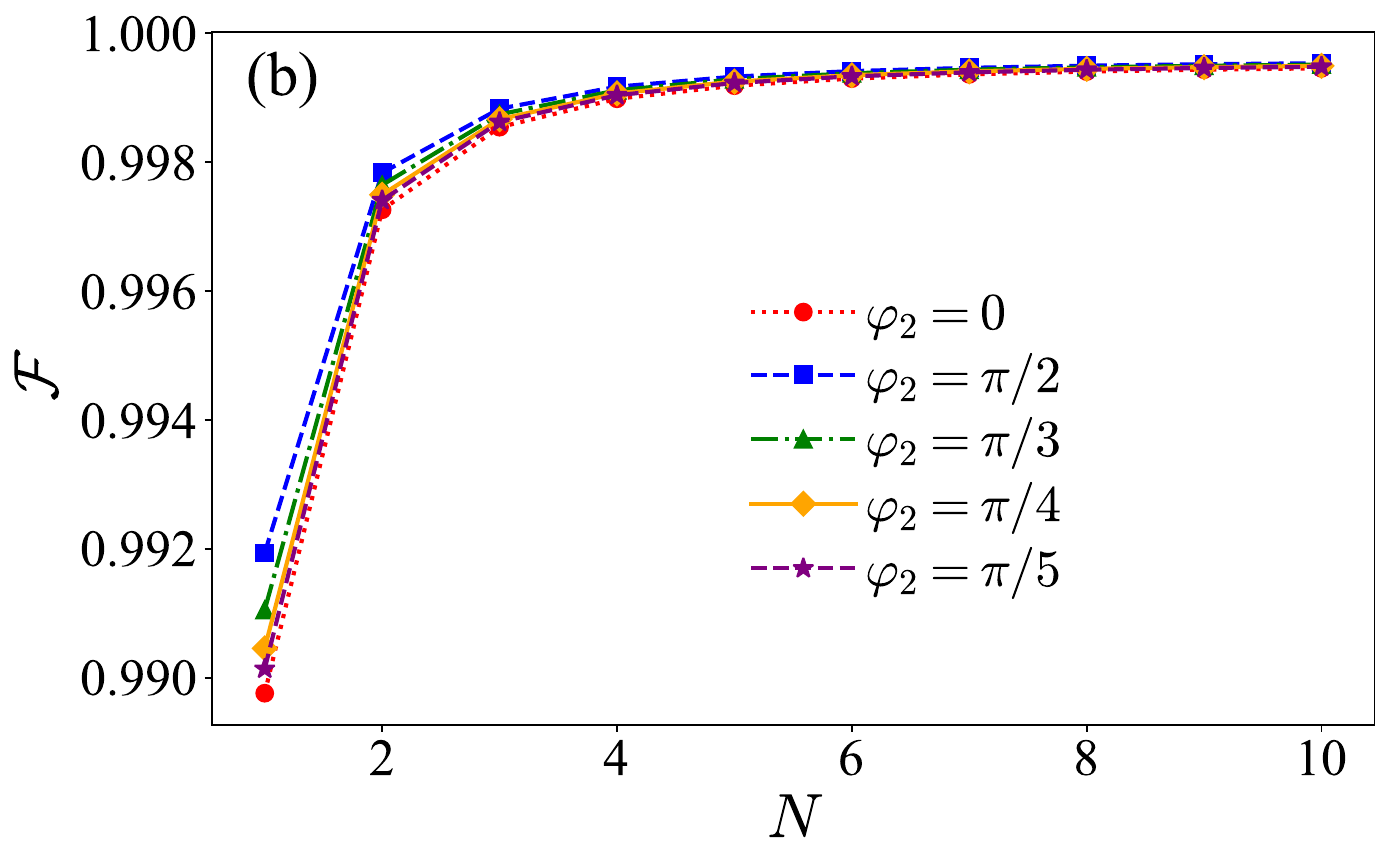}} 
	\end{minipage}
	\caption{(a) Fidelity curves of the fSim gate under the optimal parameter pulse scheme as a function of $\delta E_z$ for the initial states with different relative phases $\varphi_1$ of, where $\varphi_2$ and $\varphi_3$ are fixed at 0. (b) Fidelity curves of the fSim gate under the optimal parameter pulse scheme as a function of $\delta E_z$ for the initial states with different relative phases $\varphi_2$ of, where $\varphi_1$ and $\varphi_3$ are fixed at 0. Here $\delta E_z = 2N\pi/T \approx 2N\pi \times 20$ MHz, and $T\approx50$ ns. The gate fidelities are obtained by the average of simulations for 1600 different initial states, with the gate parameters set to $\vartheta=\pi/4$ and $\Xi=\pi/2$.}
	\label{F9}
\end{figure}\\
\twocolumngrid 
\section{AVERAGE FIDELITY FOR DIFFERENT RELATIVE PHASES OF THE INITIAL STATES\label{Appendix C}}
For the fidelity calculation in the main text, we define the average fidelity as given in Eq. (\ref{Eq.37}), and choose the initial state as the separable pure state $|\psi_0\rangle=(\cos\varPhi_1|0\rangle_1+\sin\varPhi_1|1\rangle_1) (\cos\varPhi_2|0\rangle_2+\sin\varPhi_2|1\rangle_2) $, where all relative phases of this initial state are set to 0. For the more general case, the initial state can be expressed as: $|\psi_0\rangle'=\cos\varPhi_1\cos\varPhi_2|00\rangle+ \cos\varPhi_1\sin\varPhi_2 e^{i\varphi_1}|01\rangle+\sin\varPhi_1\cos\varPhi_2e^{i\varphi_2}|10\rangle+\sin\varPhi_1\sin\varPhi_2e^{i\varphi_3}|11\rangle$, where $\varphi_{1,2,3}$ denote the relative phases between $|01\rangle$, $|10\rangle$, $|11\rangle$ and $|00\rangle$, respectively.\\
\indent Here, we investigate the influence of different phase selections of the initial state on the fidelities. To control the time cost of simulation calculations, we sequentially adjust one of the phases $\varphi_{1,2,3}$
to a set of fixed values, while fixing the other two phases fixed at 0; this allows us to separately derive the variation trends of fidelity with respect to each relative phase.
Taking the fidelity of the fSim gate under the optimal parameter pulse scheme as an example, numerical simulations show that the fidelity of the fSim gate is independent of the variation in $\varphi_3$.
Furthermore, for different initial phases $\varphi_1$ and $\varphi_2$, the fidelity of the fSim gate exhibits the following regularities: $\mathcal{F}(\varphi_1)=\mathcal{F}(\varphi_1+\pi), \mathcal{F}(\varphi_2)=\mathcal{F}(-\varphi_1)$. These symmetries are determined by the symmetry of the Hamiltonian. We thus restrict the value ranges of $\varphi_1$ and $\varphi_2$ to $[0,\pi/2]$, respectively, and sequentially assign a series of discrete $\pi/5,\pi/4,\pi/3,...$ for the two phases.\\
\indent The fidelities of the fSim gate under the optimal parameter pulse scheme for different relative phases of the initial states are plotted in Fig. \ref{F9}. It can be seen that the fidelities exhibit no significant variation for different $\varphi_{1,2}$ phases. As $\delta E_z$ increases, the fidelity loss is dominated by decoherence, and the fidelities of initial states with distinct phase configurations all converge to the same value of 99.95\%.

\end{document}